\documentclass[aps,epsfigure,superscriptaddress,nofootinbib]{revtex4-2}
\usepackage{makeidx} % Produzir indice remissivo
\usepackage{graphicx} % Incluir graficos no arquivo.
\usepackage{dcolumn} %Permitir alinhamento decimal em tabelas. Precisa do Array Package.
\usepackage{array} % Adicionar novas particularidades para as tabelas e vetores.
\usepackage{amssymb} %Permite o uso de sofisticados simbolos matematicos
\usepackage{amsmath}
\usepackage{physics}
\usepackage{textcomp}
\usepackage{multirow}
\usepackage{subfigure}
\usepackage{eucal}
\usepackage{mathrsfs}
\usepackage[all]{xy}
\usepackage{epstopdf}
\usepackage{color}
\usepackage{float} %????
\usepackage{amsmath} %????
\usepackage{amsfonts} %????
\usepackage{indentfirst} %TAB nos parágrafos
\usepackage[breakable]{tcolorbox}
    \makeatletter
    \newcommand{\boxspacing}{\kern\kvtcb@left@rule\kern\kvtcb@boxsep}
    \makeatother

 \definecolor{incolor}{HTML}{303F9F}
    \definecolor{outcolor}{HTML}{D84315}
    \definecolor{cellborder}{HTML}{CFCFCF}
    \definecolor{cellbackground}{HTML}{F7F7F7}
  
 \usepackage{upquote} % Upright quotes for verbatim code
    \usepackage{eurosym} % defines \euro
    \usepackage[mathletters]{ucs} % Extended unicode (utf-8) support
    \usepackage{fancyvrb} % verbatim replacement that allows latex
    \usepackage{grffile} % extends the file name processing of package graphics 
    \makeatletter % fix for grffile with XeLaTeX
    \def\Gread@@xetex#1{%
      \IfFileExists{"\Gin@base".bb}%
      {\Gread@eps{\Gin@base.bb}}%
      {\Gread@@xetex@aux#1}%
    }
    \makeatother
 
    \DefineVerbatimEnvironment{Highlighting}{Verbatim}{commandchars=\\\{\}}

    \DefineVerbatimEnvironment{Highlighting}{Verbatim}{commandchars=\\\{\}}
    \let\Oldtex\TeX
    \let\Oldlatex\LaTeX
    \renewcommand{\TeX}{\textrm{\Oldtex}}
    \renewcommand{\LaTeX}{\textrm{\Oldlatex}}
    \title{artigo_qiskit}

\makeatletter
\def\PY@reset{\let\PY@it=\relax \let\PY@bf=\relax%
    \let\PY@ul=\relax \let\PY@tc=\relax%
    \let\PY@bc=\relax \let\PY@ff=\relax}
\def\PY@tok#1{\csname PY@tok@#1\endcsname}
\def\PY@toks#1+{\ifx\relax#1\empty\else%
    \PY@tok{#1}\expandafter\PY@toks\fi}
\def\PY@do#1{\PY@bc{\PY@tc{\PY@ul{%
    \PY@it{\PY@bf{\PY@ff{#1}}}}}}}
\def\PY#1#2{\PY@reset\PY@toks#1+\relax+\PY@do{#2}}

\expandafter\def\csname PY@tok@w\endcsname{\def\PY@tc##1{\textcolor[rgb]{0.73,0.73,0.73}{##1}}}
\expandafter\def\csname PY@tok@c\endcsname{\let\PY@it=\textit\def\PY@tc##1{\textcolor[rgb]{0.25,0.50,0.50}{##1}}}
\expandafter\def\csname PY@tok@cp\endcsname{\def\PY@tc##1{\textcolor[rgb]{0.74,0.48,0.00}{##1}}}
\expandafter\def\csname PY@tok@k\endcsname{\let\PY@bf=\textbf\def\PY@tc##1{\textcolor[rgb]{0.00,0.50,0.00}{##1}}}
\expandafter\def\csname PY@tok@kp\endcsname{\def\PY@tc##1{\textcolor[rgb]{0.00,0.50,0.00}{##1}}}
\expandafter\def\csname PY@tok@kt\endcsname{\def\PY@tc##1{\textcolor[rgb]{0.69,0.00,0.25}{##1}}}
\expandafter\def\csname PY@tok@o\endcsname{\def\PY@tc##1{\textcolor[rgb]{0.40,0.40,0.40}{##1}}}
\expandafter\def\csname PY@tok@ow\endcsname{\let\PY@bf=\textbf\def\PY@tc##1{\textcolor[rgb]{0.67,0.13,1.00}{##1}}}
\expandafter\def\csname PY@tok@nb\endcsname{\def\PY@tc##1{\textcolor[rgb]{0.00,0.50,0.00}{##1}}}
\expandafter\def\csname PY@tok@nf\endcsname{\def\PY@tc##1{\textcolor[rgb]{0.00,0.00,1.00}{##1}}}
\expandafter\def\csname PY@tok@nc\endcsname{\let\PY@bf=\textbf\def\PY@tc##1{\textcolor[rgb]{0.00,0.00,1.00}{##1}}}
\expandafter\def\csname PY@tok@nn\endcsname{\let\PY@bf=\textbf\def\PY@tc##1{\textcolor[rgb]{0.00,0.00,1.00}{##1}}}
\expandafter\def\csname PY@tok@ne\endcsname{\let\PY@bf=\textbf\def\PY@tc##1{\textcolor[rgb]{0.82,0.25,0.23}{##1}}}
\expandafter\def\csname PY@tok@nv\endcsname{\def\PY@tc##1{\textcolor[rgb]{0.10,0.09,0.49}{##1}}}
\expandafter\def\csname PY@tok@no\endcsname{\def\PY@tc##1{\textcolor[rgb]{0.53,0.00,0.00}{##1}}}
\expandafter\def\csname PY@tok@nl\endcsname{\def\PY@tc##1{\textcolor[rgb]{0.63,0.63,0.00}{##1}}}
\expandafter\def\csname PY@tok@ni\endcsname{\let\PY@bf=\textbf\def\PY@tc##1{\textcolor[rgb]{0.60,0.60,0.60}{##1}}}
\expandafter\def\csname PY@tok@na\endcsname{\def\PY@tc##1{\textcolor[rgb]{0.49,0.56,0.16}{##1}}}
\expandafter\def\csname PY@tok@nt\endcsname{\let\PY@bf=\textbf\def\PY@tc##1{\textcolor[rgb]{0.00,0.50,0.00}{##1}}}
\expandafter\def\csname PY@tok@nd\endcsname{\def\PY@tc##1{\textcolor[rgb]{0.67,0.13,1.00}{##1}}}
\expandafter\def\csname PY@tok@s\endcsname{\def\PY@tc##1{\textcolor[rgb]{0.73,0.13,0.13}{##1}}}
\expandafter\def\csname PY@tok@sd\endcsname{\let\PY@it=\textit\def\PY@tc##1{\textcolor[rgb]{0.73,0.13,0.13}{##1}}}
\expandafter\def\csname PY@tok@si\endcsname{\let\PY@bf=\textbf\def\PY@tc##1{\textcolor[rgb]{0.73,0.40,0.53}{##1}}}
\expandafter\def\csname PY@tok@se\endcsname{\let\PY@bf=\textbf\def\PY@tc##1{\textcolor[rgb]{0.73,0.40,0.13}{##1}}}
\expandafter\def\csname PY@tok@sr\endcsname{\def\PY@tc##1{\textcolor[rgb]{0.73,0.40,0.53}{##1}}}
\expandafter\def\csname PY@tok@ss\endcsname{\def\PY@tc##1{\textcolor[rgb]{0.10,0.09,0.49}{##1}}}
\expandafter\def\csname PY@tok@sx\endcsname{\def\PY@tc##1{\textcolor[rgb]{0.00,0.50,0.00}{##1}}}
\expandafter\def\csname PY@tok@m\endcsname{\def\PY@tc##1{\textcolor[rgb]{0.40,0.40,0.40}{##1}}}
\expandafter\def\csname PY@tok@gh\endcsname{\let\PY@bf=\textbf\def\PY@tc##1{\textcolor[rgb]{0.00,0.00,0.50}{##1}}}
\expandafter\def\csname PY@tok@gu\endcsname{\let\PY@bf=\textbf\def\PY@tc##1{\textcolor[rgb]{0.50,0.00,0.50}{##1}}}
\expandafter\def\csname PY@tok@gd\endcsname{\def\PY@tc##1{\textcolor[rgb]{0.63,0.00,0.00}{##1}}}
\expandafter\def\csname PY@tok@gi\endcsname{\def\PY@tc##1{\textcolor[rgb]{0.00,0.63,0.00}{##1}}}
\expandafter\def\csname PY@tok@gr\endcsname{\def\PY@tc##1{\textcolor[rgb]{1.00,0.00,0.00}{##1}}}
\expandafter\def\csname PY@tok@ge\endcsname{\let\PY@it=\textit}
\expandafter\def\csname PY@tok@gs\endcsname{\let\PY@bf=\textbf}
\expandafter\def\csname PY@tok@gp\endcsname{\let\PY@bf=\textbf\def\PY@tc##1{\textcolor[rgb]{0.00,0.00,0.50}{##1}}}
\expandafter\def\csname PY@tok@go\endcsname{\def\PY@tc##1{\textcolor[rgb]{0.53,0.53,0.53}{##1}}}
\expandafter\def\csname PY@tok@gt\endcsname{\def\PY@tc##1{\textcolor[rgb]{0.00,0.27,0.87}{##1}}}
\expandafter\def\csname PY@tok@err\endcsname{\def\PY@bc##1{\setlength{\fboxsep}{0pt}\fcolorbox[rgb]{1.00,0.00,0.00}{1,1,1}{\strut ##1}}}
\expandafter\def\csname PY@tok@kc\endcsname{\let\PY@bf=\textbf\def\PY@tc##1{\textcolor[rgb]{0.00,0.50,0.00}{##1}}}
\expandafter\def\csname PY@tok@kd\endcsname{\let\PY@bf=\textbf\def\PY@tc##1{\textcolor[rgb]{0.00,0.50,0.00}{##1}}}
\expandafter\def\csname PY@tok@kn\endcsname{\let\PY@bf=\textbf\def\PY@tc##1{\textcolor[rgb]{0.00,0.50,0.00}{##1}}}
\expandafter\def\csname PY@tok@kr\endcsname{\let\PY@bf=\textbf\def\PY@tc##1{\textcolor[rgb]{0.00,0.50,0.00}{##1}}}
\expandafter\def\csname PY@tok@bp\endcsname{\def\PY@tc##1{\textcolor[rgb]{0.00,0.50,0.00}{##1}}}
\expandafter\def\csname PY@tok@fm\endcsname{\def\PY@tc##1{\textcolor[rgb]{0.00,0.00,1.00}{##1}}}
\expandafter\def\csname PY@tok@vc\endcsname{\def\PY@tc##1{\textcolor[rgb]{0.10,0.09,0.49}{##1}}}
\expandafter\def\csname PY@tok@vg\endcsname{\def\PY@tc##1{\textcolor[rgb]{0.10,0.09,0.49}{##1}}}
\expandafter\def\csname PY@tok@vi\endcsname{\def\PY@tc##1{\textcolor[rgb]{0.10,0.09,0.49}{##1}}}
\expandafter\def\csname PY@tok@vm\endcsname{\def\PY@tc##1{\textcolor[rgb]{0.10,0.09,0.49}{##1}}}
\expandafter\def\csname PY@tok@sa\endcsname{\def\PY@tc##1{\textcolor[rgb]{0.73,0.13,0.13}{##1}}}
\expandafter\def\csname PY@tok@sb\endcsname{\def\PY@tc##1{\textcolor[rgb]{0.73,0.13,0.13}{##1}}}
\expandafter\def\csname PY@tok@sc\endcsname{\def\PY@tc##1{\textcolor[rgb]{0.73,0.13,0.13}{##1}}}
\expandafter\def\csname PY@tok@dl\endcsname{\def\PY@tc##1{\textcolor[rgb]{0.73,0.13,0.13}{##1}}}
\expandafter\def\csname PY@tok@s2\endcsname{\def\PY@tc##1{\textcolor[rgb]{0.73,0.13,0.13}{##1}}}
\expandafter\def\csname PY@tok@sh\endcsname{\def\PY@tc##1{\textcolor[rgb]{0.73,0.13,0.13}{##1}}}
\expandafter\def\csname PY@tok@s1\endcsname{\def\PY@tc##1{\textcolor[rgb]{0.73,0.13,0.13}{##1}}}
\expandafter\def\csname PY@tok@mb\endcsname{\def\PY@tc##1{\textcolor[rgb]{0.40,0.40,0.40}{##1}}}
\expandafter\def\csname PY@tok@mf\endcsname{\def\PY@tc##1{\textcolor[rgb]{0.40,0.40,0.40}{##1}}}
\expandafter\def\csname PY@tok@mh\endcsname{\def\PY@tc##1{\textcolor[rgb]{0.40,0.40,0.40}{##1}}}
\expandafter\def\csname PY@tok@mi\endcsname{\def\PY@tc##1{\textcolor[rgb]{0.40,0.40,0.40}{##1}}}
\expandafter\def\csname PY@tok@il\endcsname{\def\PY@tc##1{\textcolor[rgb]{0.40,0.40,0.40}{##1}}}
\expandafter\def\csname PY@tok@mo\endcsname{\def\PY@tc##1{\textcolor[rgb]{0.40,0.40,0.40}{##1}}}
\expandafter\def\csname PY@tok@ch\endcsname{\let\PY@it=\textit\def\PY@tc##1{\textcolor[rgb]{0.25,0.50,0.50}{##1}}}
\expandafter\def\csname PY@tok@cm\endcsname{\let\PY@it=\textit\def\PY@tc##1{\textcolor[rgb]{0.25,0.50,0.50}{##1}}}
\expandafter\def\csname PY@tok@cpf\endcsname{\let\PY@it=\textit\def\PY@tc##1{\textcolor[rgb]{0.25,0.50,0.50}{##1}}}
\expandafter\def\csname PY@tok@c1\endcsname{\let\PY@it=\textit\def\PY@tc##1{\textcolor[rgb]{0.25,0.50,0.50}{##1}}}
\expandafter\def\csname PY@tok@cs\endcsname{\let\PY@it=\textit\def\PY@tc##1{\textcolor[rgb]{0.25,0.50,0.50}{##1}}}

\def\PYZus{\char`\_}
\def\PYZob{\char`\{}
\def\PYZcb{\char`\}}

\def\PYZsh{\char`\#}
\def\PYZpc{\char`\%}

\def\PYZhy{\char`\-}
\def\PYZsq{\char`\'}
\def\PYZdq{\char`\"}

\makeatother

    \makeatletter
        \newbox\Wrappedcontinuationbox 
        \newbox\Wrappedvisiblespacebox 
        \newcommand*\Wrappedvisiblespace {\textcolor{red}{\textvisiblespace}} 
        \newcommand*\Wrappedcontinuationsymbol {\textcolor{red}{\llap{\tiny$\m@th\hookrightarrow$}}} 
        \newcommand*\Wrappedcontinuationindent {3ex } 
        \newcommand*\Wrappedafterbreak {\kern\Wrappedcontinuationindent\copy\Wrappedcontinuationbox} 
        \newcommand*\Wrappedbreaksatspecials {% 
            \def\PYGZus{\discretionary{\char`\_}{\Wrappedafterbreak}{\char`\_}}% 
            \def\PYGZob{\discretionary{}{\Wrappedafterbreak\char`\{}{\char`\{}}% 
            \def\PYGZcb{\discretionary{\char`\}}{\Wrappedafterbreak}{\char`\}}}% 
            \def\PYGZca{\discretionary{\char`\^}{\Wrappedafterbreak}{\char`\^}}% 
            \def\PYGZam{\discretionary{\char`\&}{\Wrappedafterbreak}{\char`\&}}% 
            \def\PYGZlt{\discretionary{}{\Wrappedafterbreak\char`\<}{\char`\<}}% 
            \def\PYGZgt{\discretionary{\char`\>}{\Wrappedafterbreak}{\char`\>}}% 
            \def\PYGZsh{\discretionary{}{\Wrappedafterbreak\char`\#}{\char`\#}}% 
            \def\PYGZpc{\discretionary{}{\Wrappedafterbreak\char`\%}{\char`\%}}% 
            \def\PYGZdl{\discretionary{}{\Wrappedafterbreak\char`\$}{\char`\$}}% 
            \def\PYGZhy{\discretionary{\char`\-}{\Wrappedafterbreak}{\char`\-}}% 
            \def\PYGZsq{\discretionary{}{\Wrappedafterbreak\textquotesingle}{\textquotesingle}}% 
            \def\PYGZdq{\discretionary{}{\Wrappedafterbreak\char`\"}{\char`\"}}% 
            \def\PYGZti{\discretionary{\char`\~}{\Wrappedafterbreak}{\char`\~}}% 
        } 
        \newcommand*\Wrappedbreaksatpunct {% 
            \lccode`\~`\.\lowercase{\def~}{\discretionary{\hbox{\char`\.}}{\Wrappedafterbreak}{\hbox{\char`\.}}}% 
            \lccode`\~`\,\lowercase{\def~}{\discretionary{\hbox{\char`\,}}{\Wrappedafterbreak}{\hbox{\char`\,}}}% 
            \lccode`\~`\;\lowercase{\def~}{\discretionary{\hbox{\char`\;}}{\Wrappedafterbreak}{\hbox{\char`\;}}}% 
            \lccode`\~`\:\lowercase{\def~}{\discretionary{\hbox{\char`\:}}{\Wrappedafterbreak}{\hbox{\char`\:}}}% 
            \lccode`\~`\?\lowercase{\def~}{\discretionary{\hbox{\char`\?}}{\Wrappedafterbreak}{\hbox{\char`\?}}}% 
            \lccode`\~`\!\lowercase{\def~}{\discretionary{\hbox{\char`\!}}{\Wrappedafterbreak}{\hbox{\char`\!}}}% 
            \lccode`\~`\/\lowercase{\def~}{\discretionary{\hbox{\char`\/}}{\Wrappedafterbreak}{\hbox{\char`\/}}}% 
            \catcode`\.\active
            \catcode`\,\active 
            \catcode`\;\active
            \catcode`\:\active
            \catcode`\?\active
            \catcode`\!\active
            \catcode`\/\active 
            \lccode`\~`\~ 	
        }
    \makeatother

    \let\OriginalVerbatim=\Verbatim
    \makeatletter
    \renewcommand{\Verbatim}[1][1]{%
        \sbox\Wrappedcontinuationbox {\Wrappedcontinuationsymbol}%
        \sbox\Wrappedvisiblespacebox {\FV@SetupFont\Wrappedvisiblespace}%
        \def\FancyVerbFormatLine ##1{\hsize\linewidth
            \vtop{\raggedright\hyphenpenalty\z@\exhyphenpenalty\z@
                \doublehyphendemerits\z@\finalhyphendemerits\z@
                \strut ##1\strut}%
        }%
        \def\FV@Space {%
            \nobreak\hskip\z@ plus\fontdimen3\font minus\fontdimen4\font
            \discretionary{\copy\Wrappedvisiblespacebox}{\Wrappedafterbreak}
            {\kern\fontdimen2\font}%
        }%
        
        \Wrappedbreaksatspecials
        \OriginalVerbatim[#1,codes*=\Wrappedbreaksatpunct]%
    }
    \makeatother

\begin{document}

\title{Effect of pure dephasing quantum noise in the quantum search algorithm: an undergraduate approach using Atos Quantum Assembly}
%\title{An example of quantum search software: implementation of Grover's algorithm using Atos Quantum Assembly}
%  \small Quantum Computing: an undergraduate approach using Qiskit}

\author{Maria Heloísa Fraga da Silva}
\affiliation{Grupo de Informação Quântica e Física Estatística, Centro de Ciências Exatas e das Tecnologias, Universidade Federal do Oeste da Bahia—Campus Reitor Edgard Santos. Rua Bertioga, 892, Morada Nobre I, 47810-059 Barreiras, Bahia, Brazil.}
\affiliation{Latin American Quantum Computing Center, High Performance Computing Center, SENAI CIMATEC. Av. Orlando Gomes, 1845, Piatã, 41650-010 Salvador, Bahia, Brazil.}
\author{Gleydson Fernandes de Jesus}
\affiliation{Latin American Quantum Computing Center, High Performance Computing Center, SENAI CIMATEC. Av. Orlando Gomes, 1845, Piatã, 41650-010 Salvador, Bahia, Brazil.}
\author{Clebson Cruz}
\email{clebson.cruz@ufob.edu.br}
\affiliation{Grupo de Informação Quântica e Física Estatística, Centro de Ciências Exatas e das Tecnologias, Universidade Federal do Oeste da Bahia—Campus Reitor Edgard Santos. Rua Bertioga, 892, Morada Nobre I, 47810-059 Barreiras, Bahia, Brazil.}

\date{\today}

\begin{abstract}
Quantum computing is tipped to lead the future of global technological progress. However, the obstacles related to quantum software development are an actual challenge to overcome. In particular, there is a discrepant lack of trained and skilled workforce to implement new quantum software on the recent quantum hardware. In this scenario, this work presents a didactic implementation of the quantum search algorithm in Atos Quantum Assembly Language (AQASM). In addition, we present the creation of a virtual quantum processor whose configurable architecture allows the analysis of induced quantum noise effects on the quantum algorithms. The codes are available throughout the manuscript so that readers, even those with little scientific computing experience, can replicate them and apply the methods discussed in this article to solve their own quantum computing projects. The presented results are consistent with theoretical predictions and demonstrate that AQASM and myQLM are powerful tools for teaching quantum computing and building, implementing, and simulating quantum hardware.

\end{abstract}

\maketitle

\section{Introduction}

The advent of quantum computing is one of the bases of the so-called second quantum revolution~\cite{maia2019experimento,terhal2018quantum,harrow2017quantum}.
The scientific community ceased to be mere observers of the laws of quantum mechanics and is now actively acting to use them in the development of practical quantum devices \cite{PRXQuantum.1.020101,atzori2019second}.
In this scenario, developing devices that work according to the laws of quantum mechanics is one of the most ambitious goals of the current century. This revolution is expected to be responsible for the major technological breakthroughs of the 21st century, and millions of dollars are already being invested in research into the development of quantum hardware by companies and universities around the world \cite{gibney2019quantum, peterssen2020quantum}. Nowadays, there is already a huge variety of quantum programming languages associated with quantum processors from remote access by our home computers using affordable development environments \cite{vietz2021decision}. Nevertheless, the available workforce to develop quantum technologies is currently chronic and disproportionate to the emerging market demand, being an actual challenge to overcome \cite{peterssen2020quantum}.

It is widely agreed in the literature that the amount of trained and specialized professionals working in the development of quantum software in the world is not enough to keep up with the rapid pace of the predictions made about the benefits of a commercial quantum computer, which may be available within a few years \cite{gyongyosi2019survey}. The recent worldwide demand for human resources in this area justifies the term \textit{quantum illiteracy}, which has been attributed to this situation, where there are not even enough quantum education teachers \cite{peterssen2020quantum}. Faced with this paradigm, the governments of the world leaders in quantum computing are planning public investments in quantum education for young people on a national scale as a technological strategy, as in the United States, Europe, and China \cite{peterssen2020quantum, raymer2019us, aiello2021achieving, riedel2019europe, macquarrie2020emerging}.  

The main challenge pointed out for creating a skilled workforce is the lack of resources within undergraduate courses and even in high school \cite{amin2019needs,santos2017computador,rabelo2018abordagem,alves2020simulating,perry2019quantum,tappert2019experience,jesus2021computaccao,candela2015undergraduate,angara2020quantum}. The current bibliography in quantum computing presents a high technicality which turns out to be a common obstacle due to the mathematical and physical prerequisites necessary to fully understand quantum mechanics \cite{uhlig2019generating}, because historically its solid foundations were consolidated long before the emergence of the first machine that made its practical application possible. Recently, some authors who aim to achieve a better understanding of quantum mechanics for a broader audience are publishing new materials in the area \cite{billig2018quantum, perry2019quantum, sutor2019dancing}. In this context, an alternative for faster and more effective diffusion of this knowledge has been the practice of applied quantum computing courses for undergraduate and high school students \cite{jesus2021computaccao,candela2015undergraduate,angara2020quantum} once the knowledge needed for one to develop quantum applications does not need to be as in-depth as that required for developing quantum computers \cite{westfall2018teaching}. Besides, the use of technology in the teaching-learning process favors the consolidation of knowledge \cite{faria} hence students will be able to actively participate in this revolution and quantum education is needed so that they can not only create but also enjoy the results of it.

However, there are two main challenges tied to the teaching of quantum computing. The first one is that our schools are naturally rooted in the classical world, so early student preparation is highly beneficial in familiarizing them with these new concepts \cite{wing2008computational} because although simple, the laws of quantum mechanics are very counter-intuitive \cite{uhlig2019generating}. The second one is that the quantum algorithms used in the teaching process must present a clear advantage compared to the existing classical counterpart. In this scenario, Grover's quantum search algorithm appears as a useful alternative for introducing quantum computing \cite{jesus2021computaccao,castillo2019classical}. The search problem is one of the bases of classical computing, being a tangible problem to a wide audience, and the quadratic speed-up compared to its classical analogs represents a clear advantage of quantum computing \cite{leider2019quantum, nielsen2002quantum}.

In this context, this work presents a didactic implementation of the 4-qubit quantum search algorithm on the Atos Quantum Assembly Language (AQASM), using the quantum software stack \textit{my Quantum Learning Machine} (myQLM) for writing and executing the quantum algorithm through a Python interface.
The PyLinalg linear algebra simulator is used via the myQLM software stack to perform the ideal noise-free simulation using our home computer. The programming code is provided throughout the text; thereby {the reader can reproduce the results, even without scientific computing experience.
On the other hand, Grover's algorithm requires the application of multi-controlled quantum gates, which is an obstacle to implementing its 4-qubit and superior versions on some real quantum processor architectures. In this regard, the NoisyQProc simulator 
is used to build an emulation of a real quantum processor with tuneable architecture, which allows the analysis of quantum noise effects in the quantum search problem. Therefore, the presented results demonstrate the efficacy of AQASM as a useful tool for teaching quantum programming to a wide audience.

\section{Quantum search algorithm}
\label{compframe}
 
The search problem is a very common topic in classical computing. Considering an unstructured database with $N$ entries, the problem consists in determining the index of the database entry ($x$) that satisfies some predefined search criteria $x = y$, where $y$ is the searched element. In this regard, one can define a so-called response function ($R(x)$), which classically maps the database entries to \texttt{True} or \texttt{False}, usually represented by the binary digits (bits) \texttt{0} and \texttt{1}, respectively. The response function assumes the value $R(x) = 0$ if and only if the entry $x$ satisfies the search criteria ($x = y$). For this, one can use the so-called Oracle subroutine \cite{jesus2021computaccao}. The Oracle's function is to query the list until it finds the searched element. Thus, as further the desired element is in the list, the greater will be the number of queries needed to find the element. In this regard, the complexity of this problem is directly related to the number of elements in the list \cite{nielsen2002quantum,castillo2020classical,figgatt2017complete}. On average, the complexity of this problem requires $\frac{N}{2}$ queries to the list. Therefore, the complexity of the classical search problem is defined of order $\mathcal{O}\left(N\right)$.

Grover's famous quantum search algorithm works by performing searches on unstructured databases, and consists of an application that illustrates the superiority that quantum computing power can assume over its classical counterpart \cite{szablowski2021understanding, grover1997quantum,figgatt2017complete,https://doi.org/10.48550/arxiv.2207.05665}. By exploiting the superposition principle, Grover's complexity is quadratically speeding up, $\mathcal{O}\left(\sqrt{N}\right)$, in contrast to the $\mathcal{O}\left({N}\right)$ complexity of the classical problem \cite{nielsen2002quantum}. The quantum superposition principle allows the algorithm to map all the database items simultaneously \cite{castillo2019classical}, reducing the number of steps and consequently optimizing the search process \cite{nielsen2002quantum, jesus2021computaccao}. 

Considering Grover's algorithm for $n=4$ qubits, one can create a list of $N=16$ items, represented by each state of the computational basis of a 4 qubits system.
Table~\ref{tab:01} shows the list of 16 items for which we chose color names randomly arranged. Each item is associated with a number from 0 to 15, as listed in the second column. In the third column, we represent each decimal as its equivalent in binary code. Finally, as Grover's algorithm performs the search on the state vectors,
each binary entry is associated  with a state vector in the computational basis for 4 qubits, %of the same numerical components
as per the last column.
\begin{table}[h]
\caption{Conversion of items into vectors.}
\centering
\begin{tabular}{|c|c|c|c|}
\hline
\texttt{ITEM} & \texttt{DECIMAL} & \texttt{BINARY} & \texttt{VECTOR} \\ \hline
\texttt{Orange} & 0 & 0000 & $\ket{0000}$ \\ \hline
\texttt{Magenta} & 1 & 0001 & $\ket{0001}$ \\ \hline
\texttt{Yellow} & 2 & 0010 & $\ket{0010}$ \\ \hline
\texttt{Violet} & 3 & 0011 & $\ket{0011}$ \\ \hline
\texttt{Beige} & 4 & 0100 & $\ket{0100}$ \\ \hline
\texttt{Purple} & 5 & 0101 & $\ket{0101}$ \\ \hline
\texttt{White} & 6 & 0110 & $\ket{0110}$ \\ \hline
\texttt{Pink} & 7 & 0111 & $\ket{0111}$ \\ \hline
\texttt{Brown} & 8 & 1000 & $\ket{1000}$ \\ \hline
\texttt{Green} & 9 & 1001 & $\ket{1001}$ \\ \hline
\texttt{Black} & 10 & 1010 & $\ket{1010}$ \\ \hline
\texttt{Cyan} & 11 & 1011 & $\ket{1011}$ \\ \hline
\texttt{Salmon} & 12 & 1100 & $\ket{1100}$ \\ \hline
\texttt{Gray} & 13 & 1101 & $\ket{1101}$ \\ \hline
\texttt{Red} & 14 & 1110 & $\ket{1110}$ \\ \hline
\texttt{Blue} & 15 & 1111 & $\ket{1111}$ \\ \hline
\end{tabular}
\label{tab:01}
\end{table}

%The first column of Table~\ref{tab:01} contains a list of 16 items for which we chose color names randomly arranged. We then associated each item with a number from 0 to 15, as listed in the second column. In the third column we represent each decimal as its equivalent in binary code. Finally, we associated each binary with a state vector of the same numerical components, as per the last column.  The item chosen as an example will be $\ket{1111}$, which position will be found through both classical and quantum methods afterwards.

In the following, we present the four steps that characterize Grover's algorithm: Initialization, Oracle, Amplitude Amplification, and Measurement \cite{jesus2021computaccao}, along with the integral code written using AQASM language and myQLM quantum software stack.
The first step is to import the computational tools needed to implement the quantum algorithm in AQASM. Box 1 shows the command cell that imports the computational tools required to implement Grover's algorithm in AQASM.
\begin{tcolorbox}[breakable, size=fbox, boxrule=1pt, pad at break*=1mm,colback=cellbackground, colframe=cellborder,coltitle=black,title=Box 1: Importing the computational tools]
\begin{Verbatim}[commandchars=\\\{\}]
\PY{k+kn}{from} \PY{n+nn}{qat}\PY{n+nn}{.}\PY{n+nn}{lang}\PY{n+nn}{.}\PY{n+nn}{AQASM} \PY{k+kn}{import} \PY{n}{Program}\PY{p}{,} \PY{n}{CNOT}\PY{p}{,} \PY{n}{H}\PY{p}{,} \PY{n}{X}\PY{p}{,} \PY{n}{QRoutine}
\end{Verbatim}
\end{tcolorbox}

Then, we allocate the amount of quantum and classical bits that will compose our quantum circuit, as shown in Box 2. The same quantity is defined for both since the bits will store the results of the measurements performed on the qubits to identify their final states \cite{jesus2021computaccao}.
\begin{tcolorbox}[breakable, size=fbox, boxrule=1pt, pad at break*=1mm,colback=cellbackground, colframe=cellborder,coltitle=black,title=Box 2: Allocating qubits and classical bits registers]
\begin{Verbatim}[commandchars=\\\{\}]
\PY{c+c1}{\PYZsh{} Creating Grover\PYZsq{}s program}
\PY{n}{grover} \PY{o}{=} \PY{n}{Program}\PY{p}{(}\PY{p}{)}

\PY{c+c1}{\PYZsh{} Allocating qubits and classical bits registers}
\PY{n}{Qr} \PY{o}{=} \PY{n}{grover}\PY{o}{.}\PY{n}{qalloc}\PY{p}{(}\PY{l+m+mi}{4}\PY{p}{)}
\PY{n}{Cr} \PY{o}{=} \PY{n}{grover}\PY{o}{.}\PY{n}{calloc}\PY{p}{(}\PY{l+m+mi}{4}\PY{p}{)}
\end{Verbatim}
\end{tcolorbox}

\subsection{Initialization}

In order to initialize the qubits in a balanced superposition represented by the state $\vert S\rangle$, one can apply the Hadamard gate on all the qubits
\begin{equation}
    \vert S\rangle = H^{\otimes 4}\vert 0 0 0 0\rangle~.
\end{equation}
\par In terms of coding, Box 3 shows the initialization process of Grover's algorithm.
\begin{tcolorbox}[breakable, size=fbox, boxrule=1pt, pad at break*=1mm,colback=cellbackground, colframe=cellborder,coltitle=black,title=Box 3: Initializing the qubits in a balanced superposition]
\begin{Verbatim}[commandchars=\\\{\}]
\PY{k}{for} \PY{n}{i} \PY{o+ow}{in} \PY{n+nb}{range}\PY{p}{(}\PY{l+m+mi}{0}\PY{p}{,}\PY{l+m+mi}{4}\PY{p}{)}\PY{p}{:}
    \PY{n}{grover}\PY{o}{.}\PY{n}{apply}\PY{p}{(}\PY{n}{H}\PY{p}{,} \PY{n}{Qr}\PY{p}{[}\PY{n}{i}\PY{p}{]}\PY{p}{)}
\end{Verbatim}
\end{tcolorbox}

\subsection{Oracle}

In the following, we must build the Oracle that searches for the desired item. 
There are two main methods used in literature to build the Oracle subroutine: the Boolean and phase inversion methods \cite{nielsen2002quantum,figgatt2017complete,jesus2021computaccao}. In the boolean technique,  the presence of an auxiliary qubit, also known as ancilla, initialized in the $\vert{1}\rangle$ state, is necessary. In this scenario, the ancilla is changed only if the input to the circuit is the sought state. However, this method is analogous to the classical search problem \cite{nielsen2002quantum,figgatt2017complete} and is generally applied to compare the computing power of the quantum superposition principle for quantum computing \cite{figgatt2017complete}.

Thus, for the sake of simplicity, and as this work aims to show the application of quantum algorithms as a tool for teaching quantum computing using the AQASM language, we opted for the most straightforward method, the phase inversion method \cite{nielsen2002quantum, figgatt2017complete}. This method excludes the need for an ancilla, and the Oracle's function in this process becomes to identify the sought element in the balanced superposition of the states of the computational base described above and to add to it a negative phase. In this context, the Oracle function can be represented by the unitary operator
\begin{eqnarray}
O \vert x\rangle =
\begin{cases} 
  -\vert x\rangle & \text{se } x = y, \\
  \ \ \vert x\rangle & \text{se } x \ne y,
\end{cases} 
\label{oraculo}
\end{eqnarray} 
where $\vert x\rangle$ is a state of the computational basis. Therefore, $O$ can be defined as a diagonal matrix, which adds a negative phase to the state corresponding to the searched item $\vert y\rangle$. It is important to highlight that there is an oracle to search for each state vector on the 4-qubit computational basis. Table \ref{tab:02} shows the quantum circuits for all oracles in the 4-qubit Grover's algorithm. 
\begin{table}[h]
\caption{Quantum circuit for the Oracle subroutines for each state vector in the 4-qubit computational basis.}
\centering
\begin{tabular}{|c|c||c|c||c|c||c|c|} 
\hline
\texttt{STATE} & \texttt{ORACLE} & \texttt{STATE} & \texttt{ORACLE} & \texttt{STATE} & \texttt{ORACLE} & \texttt{STATE} & \texttt{ORACLE} \\ \hline
$\ket{Orange}$ & \includegraphics[scale=0.33]{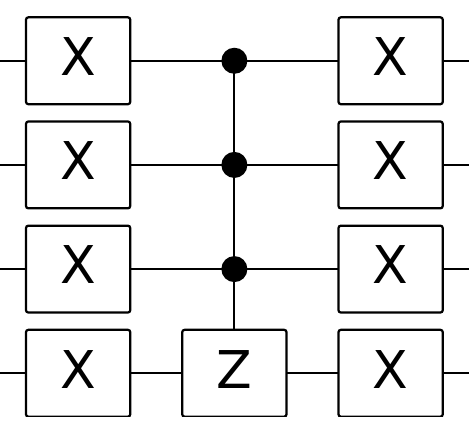} & $\ket{Beige}$ & \includegraphics[scale=0.33]{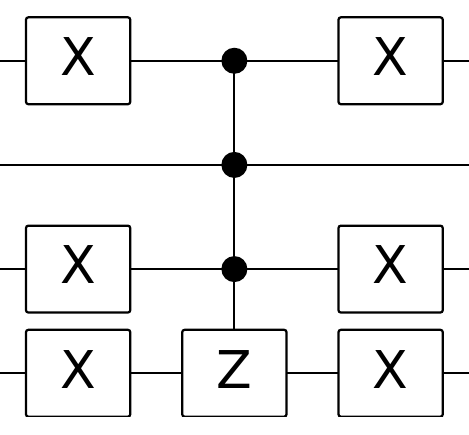} & $\ket{Brown}$ & \includegraphics[scale=0.33]{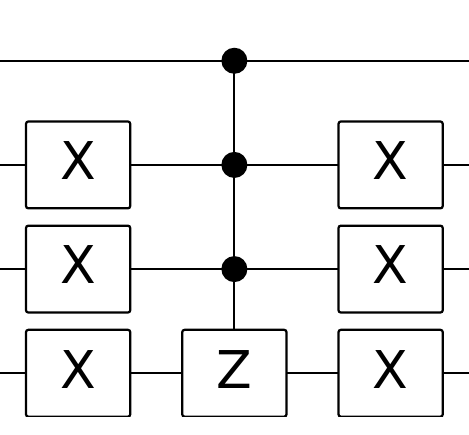} & $\ket{Salmon}$ & \includegraphics[scale=0.33]{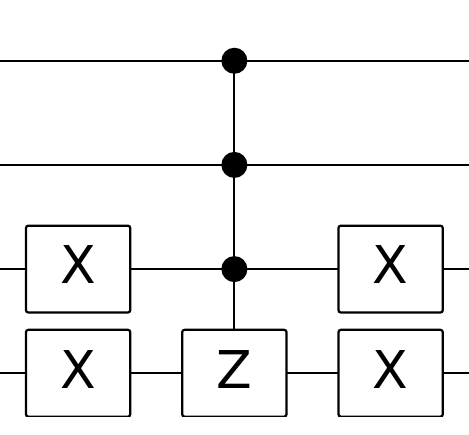} \\ \hline

$\ket{Magenta}$ & \includegraphics[scale=0.33]{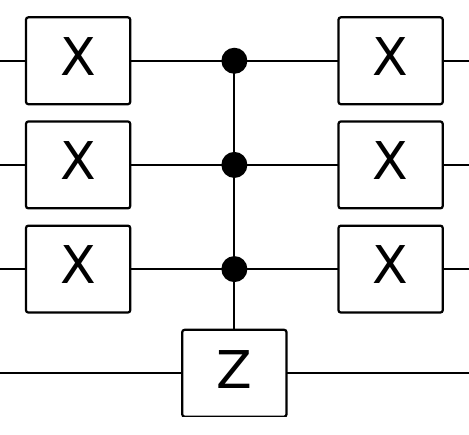} & $\ket{Purple}$ & \includegraphics[scale=0.33]{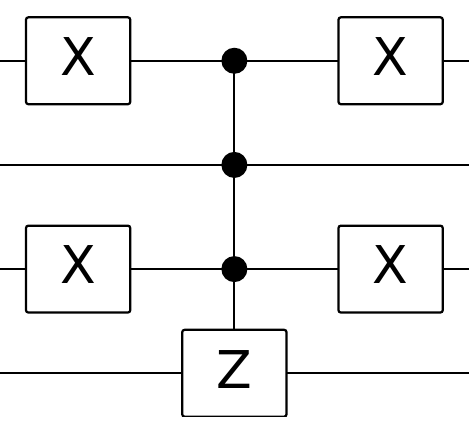} & $\ket{Green}$& \vspace{2pt} \includegraphics[scale=0.33]{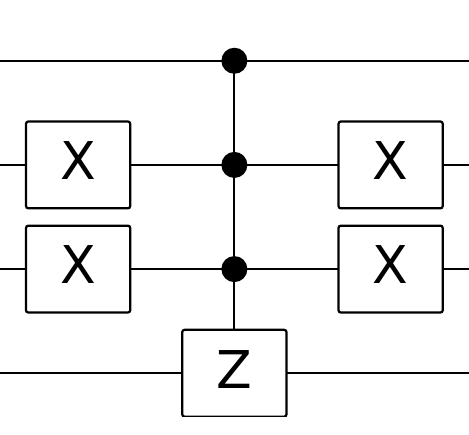} & $\ket{Gray}$ & \includegraphics[scale=0.33]{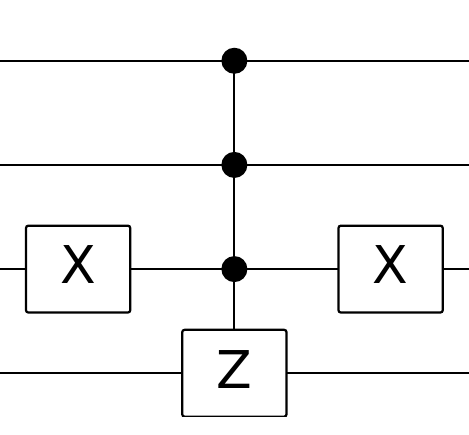} \\ \hline

$\ket{Yellow}$ & \includegraphics[scale=0.33]{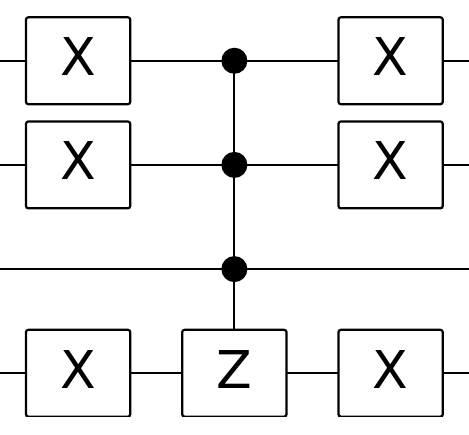} & $\ket{White}$ & \includegraphics[scale=0.33]{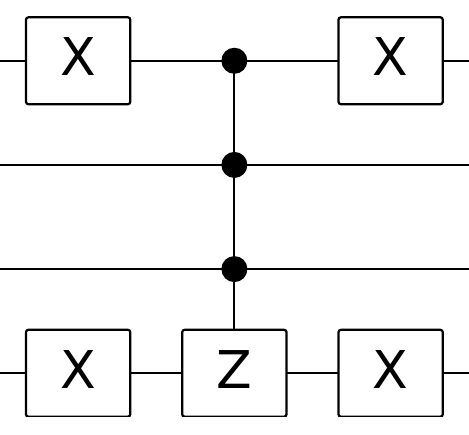} & $\ket{Black}$ & \includegraphics[scale=0.33]{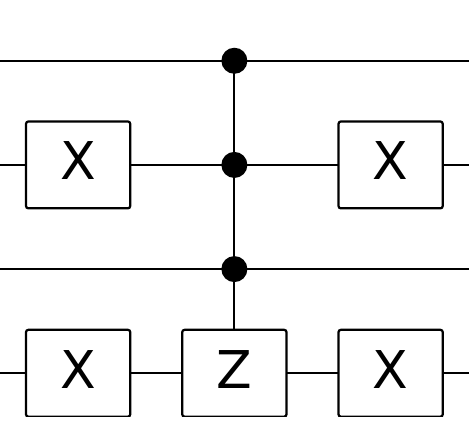} & $\ket{Red}$ & \includegraphics[scale=0.33]{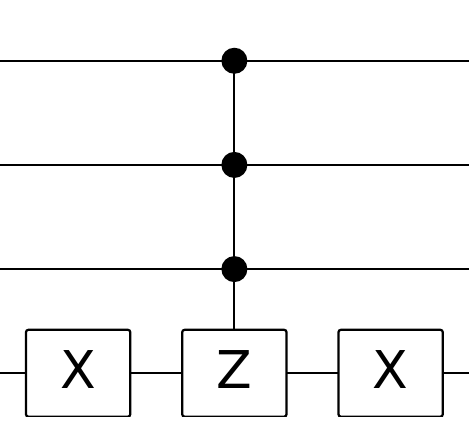} \\ \hline

$\ket{Violet}$ & \includegraphics[scale=0.33]{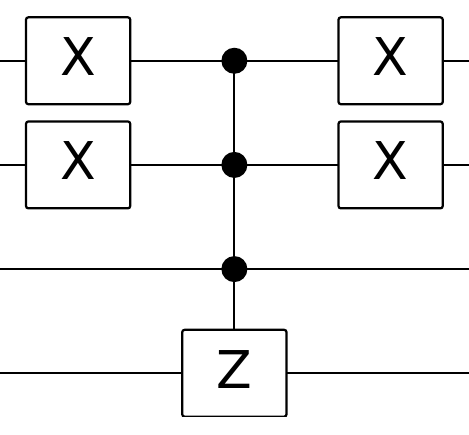} & $\ket{Pink}$ & \includegraphics[scale=0.33]{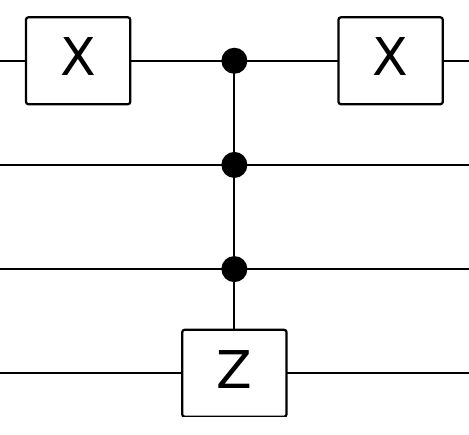} & $\ket{Cyan}$ & \includegraphics[scale=0.33]{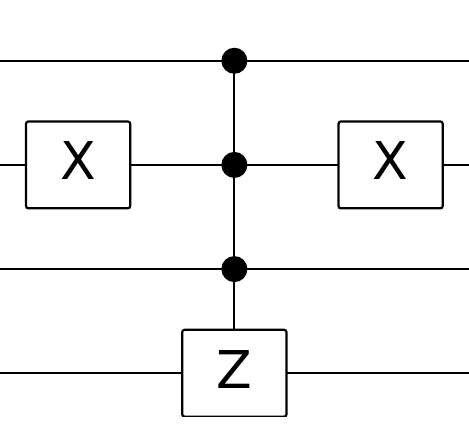} & $\ket{Blue}$ & \includegraphics[scale=0.33]{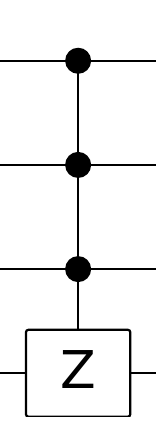} \\ \hline

\hline
\end{tabular}
\label{tab:02}
\end{table}

Here we arbitrarily choose to find the item ``Blue", associated with decimal 15, which correspondent binary is 1111 and state vector $\ket{1111}$ according to Table ~\ref{tab:01}. The oracle responsible for marking this item is composed solely of the multi-controlled gate Z (MCZ), having qubits 0 to 3 as the controls and qubit 4 as the target, as shown in Box 4. 

\begin{tcolorbox}[breakable, size=fbox, boxrule=1pt, pad at break*=1mm,colback=cellbackground, colframe=cellborder,coltitle=black,title=Box 4: Creating the Oracle subroutine that marks state $\ket{1111}$]
\begin{Verbatim}[commandchars=\\\{\}]
\PY{n}{oracle\PYZus{}1111} \PY{o}{=} \PY{n}{QRoutine}\PY{p}{(}\PY{p}{)}

\PY{c+c1}{\PYZsh{} Building the multi\PYZhy{}qubit controlled Z gate}
\PY{k}{def} \PY{n+nf}{MCZ}\PY{p}{(}\PY{n}{nb\PYZus{}control}\PY{p}{)}\PY{p}{:} 
    \PY{k}{return} \PY{n}{Z}\PY{o}{.}\PY{n}{ctrl}\PY{p}{(}\PY{n}{nb\PYZus{}control}\PY{o}{\PYZhy{}}\PY{l+m+mi}{1}\PY{p}{)}

\PY{c+c1}{\PYZsh{} Applying the multi\PYZhy{}qubit controlled Z gate}
\PY{n}{oracle\PYZus{}1111}\PY{o}{.}\PY{n}{apply}\PY{p}{(}\PY{n}{MCZ}\PY{p}{(}\PY{l+m+mi}{4}\PY{p}{)}\PY{p}{,} \PY{n}{Qr}\PY{p}{[}\PY{l+m+mi}{0}\PY{p}{]}\PY{p}{,} \PY{n}{Qr}\PY{p}{[}\PY{l+m+mi}{1}\PY{p}{]}\PY{p}{,} \PY{n}{Qr}\PY{p}{[}\PY{l+m+mi}{2}\PY{p}{]}\PY{p}{,} \PY{n}{Qr}\PY{p}{[}\PY{l+m+mi}{3}\PY{p}{]}\PY{p}{)}
\end{Verbatim}
\end{tcolorbox}

%Before, we can define the two auxiliary subroutines, Oracle and Amplification through the \textit{QRoutine} command as shown in Boxes 3 and 4, respectively.
However, even having indicated the sought element with a negative phase, the Oracle routine is still insufficient to guarantee that the searched element in the list will be found if we measure our balanced superposition. This is due to the fact that adding the phase in the sought state does not affect the probability distribution of the global state. In this sense, the probability of finding the searched item is $1/N$ ($6.25\%$), which is equivalent to the classical Oracle in a single query in the list. Therefore, it is necessary to amplify the probability of the sought element, increasing the chance of finding it and reducing the probabilities of the other states in the process.
This step is performed by the Amplitude Amplification method \cite{nielsen2002quantum,castillo2019classical,figgatt2017complete}.

\subsection{Amplitude Amplification}

In contrast to the Oracle, the Amplitude Amplification subroutine is the same, regardless of the state representing the item sought. The amplification is done by performing a reflection represented by the unitary operation 
\begin{equation}
    A = 2 \vert S\rangle\langle S \vert - \mathbb{I}~,
\end{equation}
increasing the amplitude of the searched item $\vert y\rangle$ \cite{figgatt2017complete}. Box 4 shows the cell that creates the Amplitude Amplification subroutine. First, we apply the Hadamard gate to all qubits in the Oracle-modified state. Then gate X is used to flip all qubits and we apply the MCZ gate. Finally, the process is finished by flipping all qubits again and applying the Hadamard gate, as shown in Box 5, obtaining the final state and amplifying the probability of the sought item being found.

\begin{tcolorbox}[breakable, size=fbox, boxrule=1pt, pad at break*=1mm,colback=cellbackground, colframe=cellborder,coltitle=black,title=Box 5: Creating the Amplitude Amplification subroutine]
\begin{Verbatim}[commandchars=\\\{\}]
\PY{n}{ampl} \PY{o}{=} \PY{n}{QRoutine}\PY{p}{(}\PY{p}{)}

\PY{k}{for} \PY{n}{i} \PY{o+ow}{in} \PY{n+nb}{range}\PY{p}{(}\PY{l+m+mi}{0}\PY{p}{,}\PY{l+m+mi}{4}\PY{p}{)}\PY{p}{:}
    \PY{n}{ampl}\PY{o}{.}\PY{n}{apply}\PY{p}{(}\PY{n}{H}\PY{p}{,} \PY{n}{Qr}\PY{p}{[}\PY{n}{i}\PY{p}{]}\PY{p}{)}
    
\PY{k}{for} \PY{n}{i} \PY{o+ow}{in} \PY{n+nb}{range}\PY{p}{(}\PY{l+m+mi}{0}\PY{p}{,}\PY{l+m+mi}{4}\PY{p}{)}\PY{p}{:}
    \PY{n}{ampl}\PY{o}{.}\PY{n}{apply}\PY{p}{(}\PY{n}{X}\PY{p}{,} \PY{n}{Qr}\PY{p}{[}\PY{n}{i}\PY{p}{]}\PY{p}{)}

\PY{c+c1}{\PYZsh{} Applying the multi\PYZhy{}qubit controlled Z gate}
\PY{n}{ampl}\PY{o}{.}\PY{n}{apply}\PY{p}{(}\PY{n}{MCZ}\PY{p}{(}\PY{l+m+mi}{4}\PY{p}{)}\PY{p}{,} \PY{n}{Qr}\PY{p}{[}\PY{l+m+mi}{0}\PY{p}{]}\PY{p}{,} \PY{n}{Qr}\PY{p}{[}\PY{l+m+mi}{1}\PY{p}{]}\PY{p}{,} \PY{n}{Qr}\PY{p}{[}\PY{l+m+mi}{2}\PY{p}{]}\PY{p}{,} \PY{n}{Qr}\PY{p}{[}\PY{l+m+mi}{3}\PY{p}{]}\PY{p}{)}

\PY{k}{for} \PY{n}{i} \PY{o+ow}{in} \PY{n+nb}{range}\PY{p}{(}\PY{l+m+mi}{0}\PY{p}{,}\PY{l+m+mi}{4}\PY{p}{)}\PY{p}{:}
    \PY{n}{ampl}\PY{o}{.}\PY{n}{apply}\PY{p}{(}\PY{n}{X}\PY{p}{,} \PY{n}{Qr}\PY{p}{[}\PY{n}{i}\PY{p}{]}\PY{p}{)}

\PY{k}{for} \PY{n}{i} \PY{o+ow}{in} \PY{n+nb}{range}\PY{p}{(}\PY{l+m+mi}{0}\PY{p}{,}\PY{l+m+mi}{4}\PY{p}{)}\PY{p}{:}
    \PY{n}{ampl}\PY{o}{.}\PY{n}{apply}\PY{p}{(}\PY{n}{H}\PY{p}{,} \PY{n}{Qr}\PY{p}{[}\PY{n}{i}\PY{p}{]}\PY{p}{)}
\end{Verbatim}
\end{tcolorbox}

When we reach the final state, Grover's algorithm is completed by measuring the amplified state. The probability of finding the searched item in a single measurement after the Amplitude Amplification process in the 4-qubit algorithmic probability is $39.0625 \%$, which is an improvement compared to the $6.25\%$, achieved in the classical algorithm in a single query of the Oracle, but still insufficient to assure that the search will be successful. 
Thus, in order to achieve the full potential of Grover's algorithm, subroutines Oracle and Amplitude Amplification need to be repeated to maximize the probability of finding the sought state in the measurement of the quantum state \cite{nielsen2002quantum}. The algorithmic probability of finding the searched item after $r$ repetitions is given by
\begin{equation}
    P=\left[\frac{2}{\sqrt{N}}\left( \frac{N-2r}{2N} +\frac{N-r}{N} \right)\right]^2
    \label{prob}.
\end{equation} 

From Eq. \eqref{prob}, it is possible to verify that  $r=\sqrt{N}$ is enough to assure that the user obtains the sought state after the measurement.
Fig.~\ref{fig:01} shows a sketch of the process, with a pictorial representation of the probability amplitudes in the 4-qubits scenario.

\begin{figure} [H]
    \centering
    \includegraphics[scale=0.35]{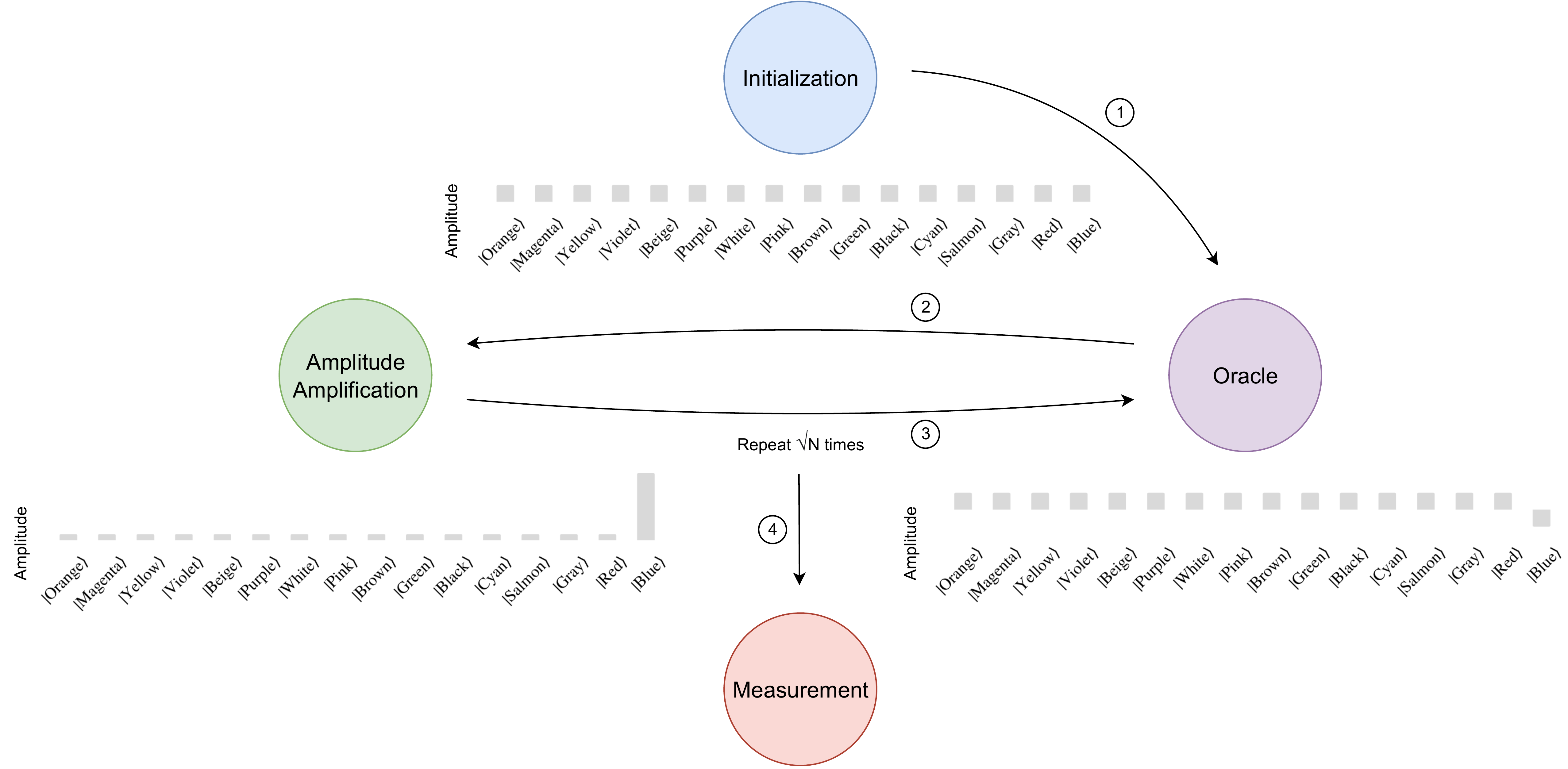}
    \caption{Sketch of the four steps of Grover’s algorithm along with the evolution of the probability amplitudes of each element of the 4-qubits computational basis.} 
    \label{fig:01}
\end{figure}

Since it is up to the user to inform which item he or she is looking for, it is possible to make available, at the beginning of the code, an interaction instructing him to type the name of the item so that our software applies the corresponding oracle (Box 6). In this context, the conditional statement \texttt{if} was used for item 0, \texttt{elif} for items 1 to 15, and \texttt{else} if the user enters an item that doesn't belong to the list. 
\begin{tcolorbox}[breakable, size=fbox, boxrule=1pt, pad at break*=1mm,colback=cellbackground, colframe=cellborder,coltitle=black,title=Box 6: Asking for user input]
\begin{Verbatim}[commandchars=\\\{\}]
\PY{n}{item} \PY{o}{=} \PY{p}{(}\PY{n+nb}{input}\PY{p}{(}\PY{l+s+s2}{\PYZdq{}}\PY{l+s+s2}{Which item do you want to find? Use initial capital letters. }\PY{l+s+s2}{\PYZdq{}}\PY{p}{)}\PY{p}{)}

\PY{k}{if} \PY{n}{item}\PY{o}{==}\PY{l+s+s2}{\PYZdq{}}\PY{l+s+s2}{Orange}\PY{l+s+s2}{\PYZdq{}}\PY{p}{:}
    \PY{c+c1}{\PYZsh{} Repeating the subroutines}
    \PY{k}{for} \PY{n}{x} \PY{o+ow}{in} \PY{n+nb}{range} \PY{p}{(}\PY{l+m+mi}{3}\PY{p}{)}\PY{p}{:}
        \PY{c+c1}{\PYZsh{} applying Oracle subroutine}
        \PY{n}{grover}\PY{o}{.}\PY{n}{apply}\PY{p}{(}\PY{n}{oracle\PYZus{}0000}\PY{p}{,}\PY{n}{Qr}\PY{p}{)}
        \PY{c+c1}{\PYZsh{} applying Amplitude Amplification subroutine}
        \PY{n}{grover}\PY{o}{.}\PY{n}{apply}\PY{p}{(}\PY{n}{ampl}\PY{p}{,}\PY{n}{Qr}\PY{p}{)}
        
\PY{k}{elif} \PY{n}{item}\PY{o}{==}\PY{l+s+s2}{\PYZdq{}}\PY{l+s+s2}{Magenta}\PY{l+s+s2}{\PYZdq{}}\PY{p}{:}
    \PY{k}{for} \PY{n}{x} \PY{o+ow}{in} \PY{n+nb}{range} \PY{p}{(}\PY{l+m+mi}{3}\PY{p}{)}\PY{p}{:}
        \PY{n}{grover}\PY{o}{.}\PY{n}{apply}\PY{p}{(}\PY{n}{oracle\PYZus{}0001}\PY{p}{,}\PY{n}{Qr}\PY{p}{)}
        \PY{n}{grover}\PY{o}{.}\PY{n}{apply}\PY{p}{(}\PY{n}{ampl}\PY{p}{,}\PY{n}{Qr}\PY{p}{)}
\PY{k}{elif} \PY{n}{item}\PY{o}{==}\PY{l+s+s2}{\PYZdq{}}\PY{l+s+s2}{Yellow}\PY{l+s+s2}{\PYZdq{}}\PY{p}{:}
    \PY{k}{for} \PY{n}{x} \PY{o+ow}{in} \PY{n+nb}{range} \PY{p}{(}\PY{l+m+mi}{3}\PY{p}{)}\PY{p}{:}
        \PY{n}{grover}\PY{o}{.}\PY{n}{apply}\PY{p}{(}\PY{n}{oracle\PYZus{}0010}\PY{p}{,}\PY{n}{Qr}\PY{p}{)}
        \PY{n}{grover}\PY{o}{.}\PY{n}{apply}\PY{p}{(}\PY{n}{ampl}\PY{p}{,}\PY{n}{Qr}\PY{p}{)}
\PY{k}{elif} \PY{n}{item}\PY{o}{==}\PY{l+s+s2}{\PYZdq{}}\PY{l+s+s2}{Violet}\PY{l+s+s2}{\PYZdq{}}\PY{p}{:}
    \PY{k}{for} \PY{n}{x} \PY{o+ow}{in} \PY{n+nb}{range} \PY{p}{(}\PY{l+m+mi}{3}\PY{p}{)}\PY{p}{:}
        \PY{n}{grover}\PY{o}{.}\PY{n}{apply}\PY{p}{(}\PY{n}{oracle\PYZus{}0011}\PY{p}{,}\PY{n}{Qr}\PY{p}{)}
        \PY{n}{grover}\PY{o}{.}\PY{n}{apply}\PY{p}{(}\PY{n}{ampl}\PY{p}{,}\PY{n}{Qr}\PY{p}{)}
\PY{k}{elif} \PY{n}{item}\PY{o}{==}\PY{l+s+s2}{\PYZdq{}}\PY{l+s+s2}{Beige}\PY{l+s+s2}{\PYZdq{}}\PY{p}{:}
    \PY{k}{for} \PY{n}{x} \PY{o+ow}{in} \PY{n+nb}{range} \PY{p}{(}\PY{l+m+mi}{3}\PY{p}{)}\PY{p}{:}
        \PY{n}{grover}\PY{o}{.}\PY{n}{apply}\PY{p}{(}\PY{n}{oracle\PYZus{}0100}\PY{p}{,}\PY{n}{Qr}\PY{p}{)}
        \PY{n}{grover}\PY{o}{.}\PY{n}{apply}\PY{p}{(}\PY{n}{ampl}\PY{p}{,}\PY{n}{Qr}\PY{p}{)}
\PY{k}{elif} \PY{n}{item}\PY{o}{==}\PY{l+s+s2}{\PYZdq{}}\PY{l+s+s2}{Purple}\PY{l+s+s2}{\PYZdq{}}\PY{p}{:}
    \PY{k}{for} \PY{n}{x} \PY{o+ow}{in} \PY{n+nb}{range} \PY{p}{(}\PY{l+m+mi}{3}\PY{p}{)}\PY{p}{:}
        \PY{n}{grover}\PY{o}{.}\PY{n}{apply}\PY{p}{(}\PY{n}{oracle\PYZus{}0101}\PY{p}{,}\PY{n}{Qr}\PY{p}{)}
        \PY{n}{grover}\PY{o}{.}\PY{n}{apply}\PY{p}{(}\PY{n}{ampl}\PY{p}{,}\PY{n}{Qr}\PY{p}{)}
\PY{k}{elif} \PY{n}{item}\PY{o}{==}\PY{l+s+s2}{\PYZdq{}}\PY{l+s+s2}{White}\PY{l+s+s2}{\PYZdq{}}\PY{p}{:}
    \PY{k}{for} \PY{n}{x} \PY{o+ow}{in} \PY{n+nb}{range} \PY{p}{(}\PY{l+m+mi}{3}\PY{p}{)}\PY{p}{:}
        \PY{n}{grover}\PY{o}{.}\PY{n}{apply}\PY{p}{(}\PY{n}{oracle\PYZus{}0110}\PY{p}{,}\PY{n}{Qr}\PY{p}{)}
        \PY{n}{grover}\PY{o}{.}\PY{n}{apply}\PY{p}{(}\PY{n}{ampl}\PY{p}{,}\PY{n}{Qr}\PY{p}{)}
\PY{k}{elif} \PY{n}{item}\PY{o}{==}\PY{l+s+s2}{\PYZdq{}}\PY{l+s+s2}{Pink}\PY{l+s+s2}{\PYZdq{}}\PY{p}{:}
    \PY{k}{for} \PY{n}{x} \PY{o+ow}{in} \PY{n+nb}{range} \PY{p}{(}\PY{l+m+mi}{3}\PY{p}{)}\PY{p}{:}
        \PY{n}{grover}\PY{o}{.}\PY{n}{apply}\PY{p}{(}\PY{n}{oracle\PYZus{}0111}\PY{p}{,}\PY{n}{Qr}\PY{p}{)}
        \PY{n}{grover}\PY{o}{.}\PY{n}{apply}\PY{p}{(}\PY{n}{ampl}\PY{p}{,}\PY{n}{Qr}\PY{p}{)}
\PY{k}{elif} \PY{n}{item}\PY{o}{==}\PY{l+s+s2}{\PYZdq{}}\PY{l+s+s2}{Brown}\PY{l+s+s2}{\PYZdq{}}\PY{p}{:}
    \PY{k}{for} \PY{n}{x} \PY{o+ow}{in} \PY{n+nb}{range} \PY{p}{(}\PY{l+m+mi}{3}\PY{p}{)}\PY{p}{:}
        \PY{n}{grover}\PY{o}{.}\PY{n}{apply}\PY{p}{(}\PY{n}{oracle\PYZus{}1000}\PY{p}{,}\PY{n}{Qr}\PY{p}{)}
        \PY{n}{grover}\PY{o}{.}\PY{n}{apply}\PY{p}{(}\PY{n}{ampl}\PY{p}{,}\PY{n}{Qr}\PY{p}{)}
\PY{k}{elif} \PY{n}{item}\PY{o}{==}\PY{l+s+s2}{\PYZdq{}}\PY{l+s+s2}{Green}\PY{l+s+s2}{\PYZdq{}}\PY{p}{:}
    \PY{k}{for} \PY{n}{x} \PY{o+ow}{in} \PY{n+nb}{range} \PY{p}{(}\PY{l+m+mi}{3}\PY{p}{)}\PY{p}{:}
        \PY{n}{grover}\PY{o}{.}\PY{n}{apply}\PY{p}{(}\PY{n}{oracle\PYZus{}1001}\PY{p}{,}\PY{n}{Qr}\PY{p}{)}
        \PY{n}{grover}\PY{o}{.}\PY{n}{apply}\PY{p}{(}\PY{n}{ampl}\PY{p}{,}\PY{n}{Qr}\PY{p}{)}
\PY{k}{elif} \PY{n}{item}\PY{o}{==}\PY{l+s+s2}{\PYZdq{}}\PY{l+s+s2}{Black}\PY{l+s+s2}{\PYZdq{}}\PY{p}{:}
    \PY{k}{for} \PY{n}{x} \PY{o+ow}{in} \PY{n+nb}{range} \PY{p}{(}\PY{l+m+mi}{3}\PY{p}{)}\PY{p}{:}
        \PY{n}{grover}\PY{o}{.}\PY{n}{apply}\PY{p}{(}\PY{n}{oracle\PYZus{}1010}\PY{p}{,}\PY{n}{Qr}\PY{p}{)}
        \PY{n}{grover}\PY{o}{.}\PY{n}{apply}\PY{p}{(}\PY{n}{ampl}\PY{p}{,}\PY{n}{Qr}\PY{p}{)}
\PY{k}{elif} \PY{n}{item}\PY{o}{==}\PY{l+s+s2}{\PYZdq{}}\PY{l+s+s2}{Cyan}\PY{l+s+s2}{\PYZdq{}}\PY{p}{:}
    \PY{k}{for} \PY{n}{x} \PY{o+ow}{in} \PY{n+nb}{range} \PY{p}{(}\PY{l+m+mi}{3}\PY{p}{)}\PY{p}{:}
        \PY{n}{grover}\PY{o}{.}\PY{n}{apply}\PY{p}{(}\PY{n}{oracle\PYZus{}1011}\PY{p}{,}\PY{n}{Qr}\PY{p}{)}
        \PY{n}{grover}\PY{o}{.}\PY{n}{apply}\PY{p}{(}\PY{n}{ampl}\PY{p}{,}\PY{n}{Qr}\PY{p}{)}
\PY{k}{elif} \PY{n}{item}\PY{o}{==}\PY{l+s+s2}{\PYZdq{}}\PY{l+s+s2}{Salmon}\PY{l+s+s2}{\PYZdq{}}\PY{p}{:}
    \PY{k}{for} \PY{n}{x} \PY{o+ow}{in} \PY{n+nb}{range} \PY{p}{(}\PY{l+m+mi}{3}\PY{p}{)}\PY{p}{:}
        \PY{n}{grover}\PY{o}{.}\PY{n}{apply}\PY{p}{(}\PY{n}{oracle\PYZus{}1100}\PY{p}{,}\PY{n}{Qr}\PY{p}{)}
        \PY{n}{grover}\PY{o}{.}\PY{n}{apply}\PY{p}{(}\PY{n}{ampl}\PY{p}{,}\PY{n}{Qr}\PY{p}{)}
\PY{k}{elif} \PY{n}{item}\PY{o}{==}\PY{l+s+s2}{\PYZdq{}}\PY{l+s+s2}{Gray}\PY{l+s+s2}{\PYZdq{}}\PY{p}{:}
    \PY{k}{for} \PY{n}{x} \PY{o+ow}{in} \PY{n+nb}{range} \PY{p}{(}\PY{l+m+mi}{3}\PY{p}{)}\PY{p}{:}
        \PY{n}{grover}\PY{o}{.}\PY{n}{apply}\PY{p}{(}\PY{n}{oracle\PYZus{}1101}\PY{p}{,}\PY{n}{Qr}\PY{p}{)}
        \PY{n}{grover}\PY{o}{.}\PY{n}{apply}\PY{p}{(}\PY{n}{ampl}\PY{p}{,}\PY{n}{Qr}\PY{p}{)}
\PY{k}{elif} \PY{n}{item}\PY{o}{==}\PY{l+s+s2}{\PYZdq{}}\PY{l+s+s2}{Red}\PY{l+s+s2}{\PYZdq{}}\PY{p}{:}
    \PY{k}{for} \PY{n}{x} \PY{o+ow}{in} \PY{n+nb}{range} \PY{p}{(}\PY{l+m+mi}{3}\PY{p}{)}\PY{p}{:}
        \PY{n}{grover}\PY{o}{.}\PY{n}{apply}\PY{p}{(}\PY{n}{oracle\PYZus{}1110}\PY{p}{,}\PY{n}{Qr}\PY{p}{)}
        \PY{n}{grover}\PY{o}{.}\PY{n}{apply}\PY{p}{(}\PY{n}{ampl}\PY{p}{,}\PY{n}{Qr}\PY{p}{)}
\PY{k}{elif} \PY{n}{item}\PY{o}{==}\PY{l+s+s2}{\PYZdq{}}\PY{l+s+s2}{Blue}\PY{l+s+s2}{\PYZdq{}}\PY{p}{:}
    \PY{k}{for} \PY{n}{x} \PY{o+ow}{in} \PY{n+nb}{range} \PY{p}{(}\PY{l+m+mi}{3}\PY{p}{)}\PY{p}{:}
        \PY{n}{grover}\PY{o}{.}\PY{n}{apply}\PY{p}{(}\PY{n}{oracle\PYZus{}1111}\PY{p}{,}\PY{n}{Qr}\PY{p}{)}
        \PY{n}{grover}\PY{o}{.}\PY{n}{apply}\PY{p}{(}\PY{n}{ampl}\PY{p}{,}\PY{n}{Qr}\PY{p}{)}
\PY{k}{else}\PY{p}{:}
    \PY{n+nb}{print}\PY{p}{(}\PY{l+s+s2}{\PYZdq{}}\PY{l+s+s2}{This item is not in the list, please check the spelling or enter another option.}\PY{l+s+s2}{\PYZdq{}}\PY{p}{)}
\end{Verbatim}
\end{tcolorbox}

At this point, the user would type the input ``Blue" implying the application of the subroutines Oracle and Amplitude Amplification.
The following output is displayed:

\texttt{
\color{blue}{Which item do you want to find? Use initial capital letters. Blue}
}

\subsection{Measurement}

The fourth and final step of the Algorithm is to perform measurements. Here we will store the final result of each qubit in the corresponding classical bit via the \textit{measure} command, as illustrated in Box 7. Note that it is necessary to transform the program into a circuit before representing it graphically in a drawing, which is laid out in SVG format as in Fig.~\ref{fig:02}.
\begin{tcolorbox}[breakable, size=fbox, boxrule=1pt, pad at break*=1mm,colback=cellbackground, colframe=cellborder,coltitle=black,title=Box 7: Measurements]
\begin{Verbatim}[commandchars=\\\{\}]
\PY{c+c1}{\PYZsh{} Measuring the qubits}
\PY{n}{grover}\PY{o}{.}\PY{n}{measure}\PY{p}{(}\PY{n}{Qr}\PY{p}{,} \PY{n}{Cr}\PY{p}{)}

\PY{c+c1}{\PYZsh{} Returning a circuit implementing Grover\PYZsq{}s program}
\PY{n}{circuit} \PY{o}{=} \PY{n}{grover}\PY{o}{.}\PY{n}{to\PYZus{}circ}\PY{p}{(}\PY{p}{)}

\PY{c+c1}{\PYZsh{} Drawing the circuit}
\PY{o}{\PYZpc{}}\PY{k}{qatdisplay} circuit  \PYZhy{}\PYZhy{}svg
\end{Verbatim}
\end{tcolorbox}

We can visualize below the complete circuit of Grover's algorithm for the Blue item search, assembled through the commands entered so far.

\begin{figure}[H]
    \centering
    \includegraphics[scale=0.5]{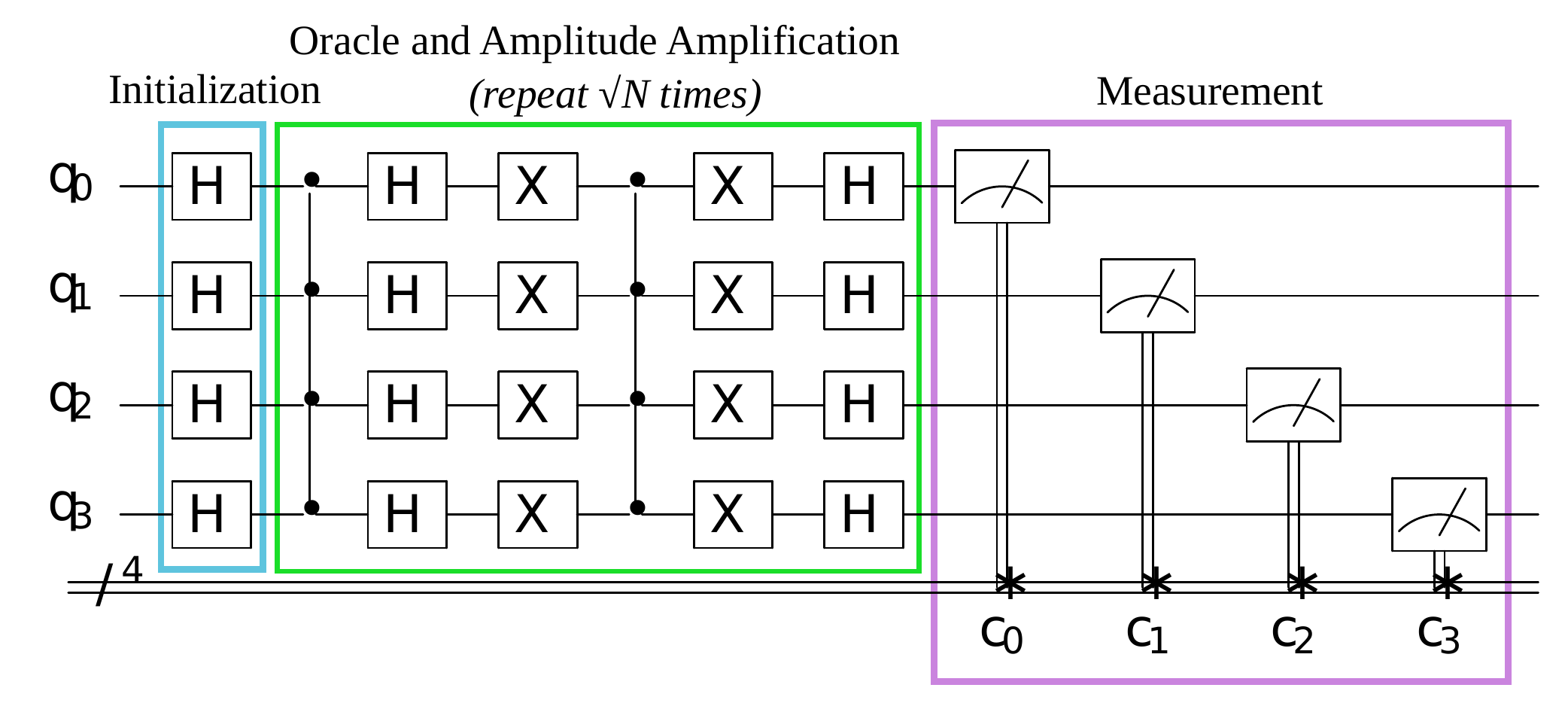}
    \caption{Grover's circuit using myQLM (abbreviated).}
    \label{fig:02}
\end{figure}

\begin{tcolorbox}[breakable, size=fbox, boxrule=1pt, pad at break*=1mm,colback=cellbackground, colframe=cellborder,coltitle=black,title=Box 8: Simulating and plotting the results]
\begin{Verbatim}[commandchars=\\\{\}]
\PY{c+c1}{\PYZsh{} Simulating the results}
\PY{k+kn}{from} \PY{n+nn}{qat}\PY{n+nn}{.}\PY{n+nn}{qpus} \PY{k+kn}{import} \PY{n}{PyLinalg}
\PY{n}{qpu} \PY{o}{=} \PY{n}{PyLinalg}\PY{p}{(}\PY{p}{)}
\PY{n}{job} \PY{o}{=} \PY{n}{circuit}\PY{o}{.}\PY{n}{to\PYZus{}job}\PY{p}{(}\PY{n}{nbshots}\PY{o}{=}\PY{l+m+mi}{8192}\PY{p}{)}
\PY{n}{result} \PY{o}{=} \PY{n}{qpu}\PY{o}{.}\PY{n}{submit}\PY{p}{(}\PY{n}{job}\PY{p}{)}

\PY{c+c1}{\PYZsh{} Printing the results}
\PY{k}{for} \PY{n}{sample} \PY{o+ow}{in} \PY{n}{result}\PY{p}{:}
    \PY{n+nb}{print}\PY{p}{(}\PY{l+s+s2}{\PYZdq{}}\PY{l+s+s2}{State }\PY{l+s+si}{\PYZpc{}s}\PY{l+s+s2}{: probability }\PY{l+s+si}{\PYZpc{}s}\PY{l+s+s2}{ +/\PYZhy{} }\PY{l+s+si}{\PYZpc{}s}\PY{l+s+s2}{\PYZdq{}} \PY{o}{\PYZpc{}} \PY{p}{(}\PY{n}{sample}\PY{o}{.}\PY{n}{state}\PY{p}{,} \PY{n}{sample}\PY{o}{.}\PY{n}{probability}\PY{p}{,} \PY{n}{sample}\PY{o}{.}\PY{n}{err}\PY{p}{)}\PY{p}{)}
\end{Verbatim}
\end{tcolorbox}

Therefore, using the quantum simulator \texttt{PyLinalg} made available by Atos and Senai CIMATEC for data processing, one can turn the search results into histogram format, as shown in Fig.~\ref{fig:03}. 
Moreover, PyLinalg simulates the execution of quantum circuits on a local processor and returns the counts of each measurement in the final state for a given set of repetitions (or shots) of that circuit in quantity defined by the user. Increasing the number of shots optimizes the exploratory simulation process, bringing it closer to the theoretically predicted result \cite{jesus2021computaccao}. As shown in Box 8, we performed $2^{13}$ shots.
\begin{figure}[H]
    \centering
    \includegraphics[scale=0.3]{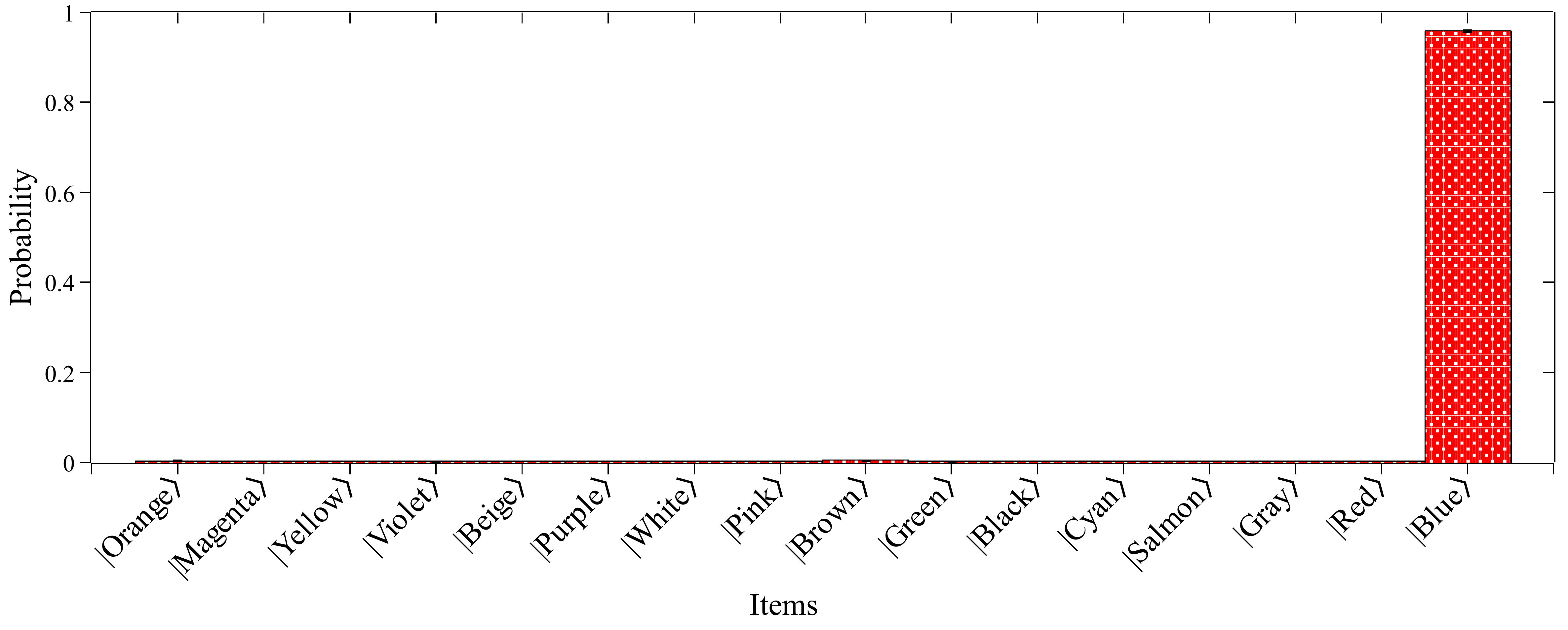}
    \caption{Probability distribution for the 16 items of the database. The searched item is found with 95.86\% probability, while the other ones did not reach 0.5\% probability.}
    \label{fig:03}
\end{figure}

As can be seen, the searched item was found with 95.86\% probability, while the other items in the list shown in Table \ref{tab:01} did not reach 0.5\% probability each. This significant increase in probability is related to the application of the repetitions of the oracle and amplitude amplification subroutines, as shown in Eq. \eqref{prob}. 
This is in contrast to the classical scenario since the probability of finding an item in an unstructured list with N=16 entries, running just one query to the list, is 6.25\%. This result highlights the benefit of using quantum features such as superposition to process information. In addition, while classically, the Oracle needs to query the list N/2 on average, the quantum algorithm can find the marked item in $\sqrt{N}$ attempts using Grover's amplitude amplification method for solving the search problem.  In conclusion, it is shown that the combination of Oracle and Amplification subroutines for developing Grover's algorithm results in a quadratic acceleration of the search problem, demonstrating that quantum computers have a major advantage over conventional computers.

\section{Quantum noise interference on Grover's algorithm}

Despite its clear advantage, Grover's algorithm requires the application of multi-controlled quantum gates \cite{jesus2021computaccao}, which is an obstacle to implementing its 4-qubit and up versions in some quantum processors architectures presents in the popular IBM Quantum Experience \cite{ibm} platform, for example. Consequently, there is a barrier to the scalability that would be required to make some algorithms useful and marketable on a large scale \cite{terhal2015quantum, preskill2018quantum}.  On the other hand, the number of qubits that are accessible in the system can be increased to substantially benefit the Grover algorithm. This would result in an exponential rise in the amount of storage space available in the database, which is the location where the searches are carried out. In this context, we carried out an emulation of a genuine quantum processor by making use of myQLM as a means of simulating the interconnectivity amongst the necessary qubits in an effort to overcome this challenge of compatibility.

In addition, the existence of noise that affects the qubits during the implementation of quantum operations or even during idle time is the primary obstacle that prevents present quantum computers from reaching their full potential. In this scenario, Grover's algorithm is one of the main hostages of noise since the quantum advantage of this algorithm can be better leveraged in quantum computers with a large number of qubits, which implies a large amount of noise and requires error correction protocols.

 It is essential to note that all measurements conducted on a system are impacted by noise to varying degrees. This is anticipated by quantum physics, which postulates that quantum states collapse instantaneously at the point of measurement. This disturbance is inherent to the theory and cannot be prevented by any measuring technique \cite{clerk2010introduction}. 
Therefore, in order to simulate the quantum noise interference of the quantum processor on Grover's algorithm, we use the quantum computing simulator KUATOMU, an ATOS QLM simulator, located at SENAI CIMATEC's HPC center in Brazil, to build a simulated quantum hardware topology, since the AQASM noisy section is restricted to QLM users.

\subsection{Quantum hardware model simulation}

The hardware model simulation used supports a maximum arity gate of $3$, thus, we decomposed the \texttt{CCCZ} gate into a set formed by \texttt{Hadamard} and \texttt{CCNOT} gates, as illustrated in Fig.~\ref{fig:04}. Consequently, it is necessary to introduce an auxiliary qubit for the circuit implementation to decompose this gate. 

\begin{figure} [H]
    \centering
    \subfigure[]{\includegraphics[scale=0.6]{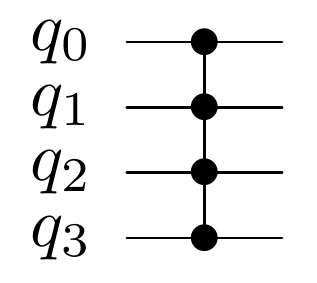}}\\
    \subfigure[]{\includegraphics[scale=0.6]{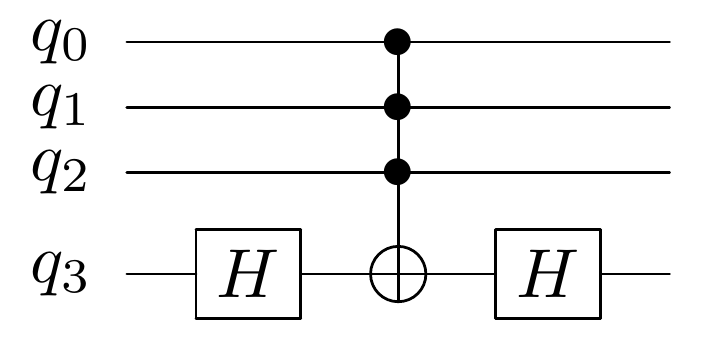}}
    \subfigure[]{\includegraphics[scale=0.6]{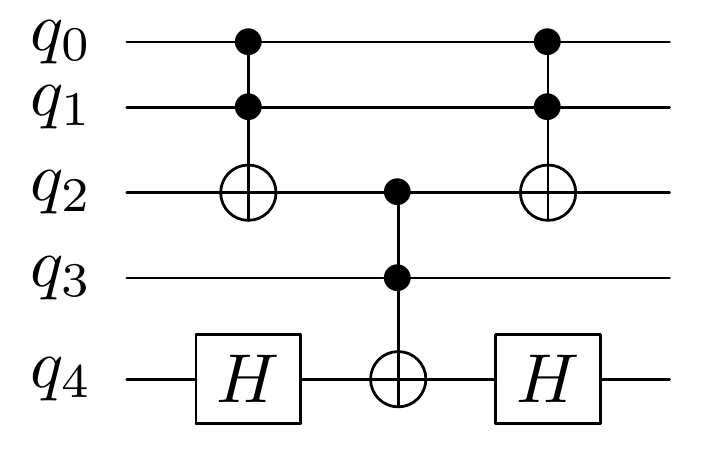}}
    \caption{The decomposition of  (a) multi-controlled Z gate ( \texttt{CCCZ}) using (b) Hadamard and CCCNOT gates or (c) Hadamard and CCNOT gates.}
    \label{fig:04}
\end{figure}

Therefore, to implement this new circuit, we created a specific 5-qubit topology in order to support the 4-qubit Grover algorithm and the auxiliary qubit of the decomposed CCCZ gate. All qubits are linked for the sake of simplicity; as a result, all of them can be concurrently controlled or targeted by controlled quantum operations. Fig.~\ref{fig:05} shows a sketch of the 5-qubit topology in which the 4-qubit Grover algorithm is implemented.

\begin{figure}[H]
\centering
\includegraphics[scale=0.27]{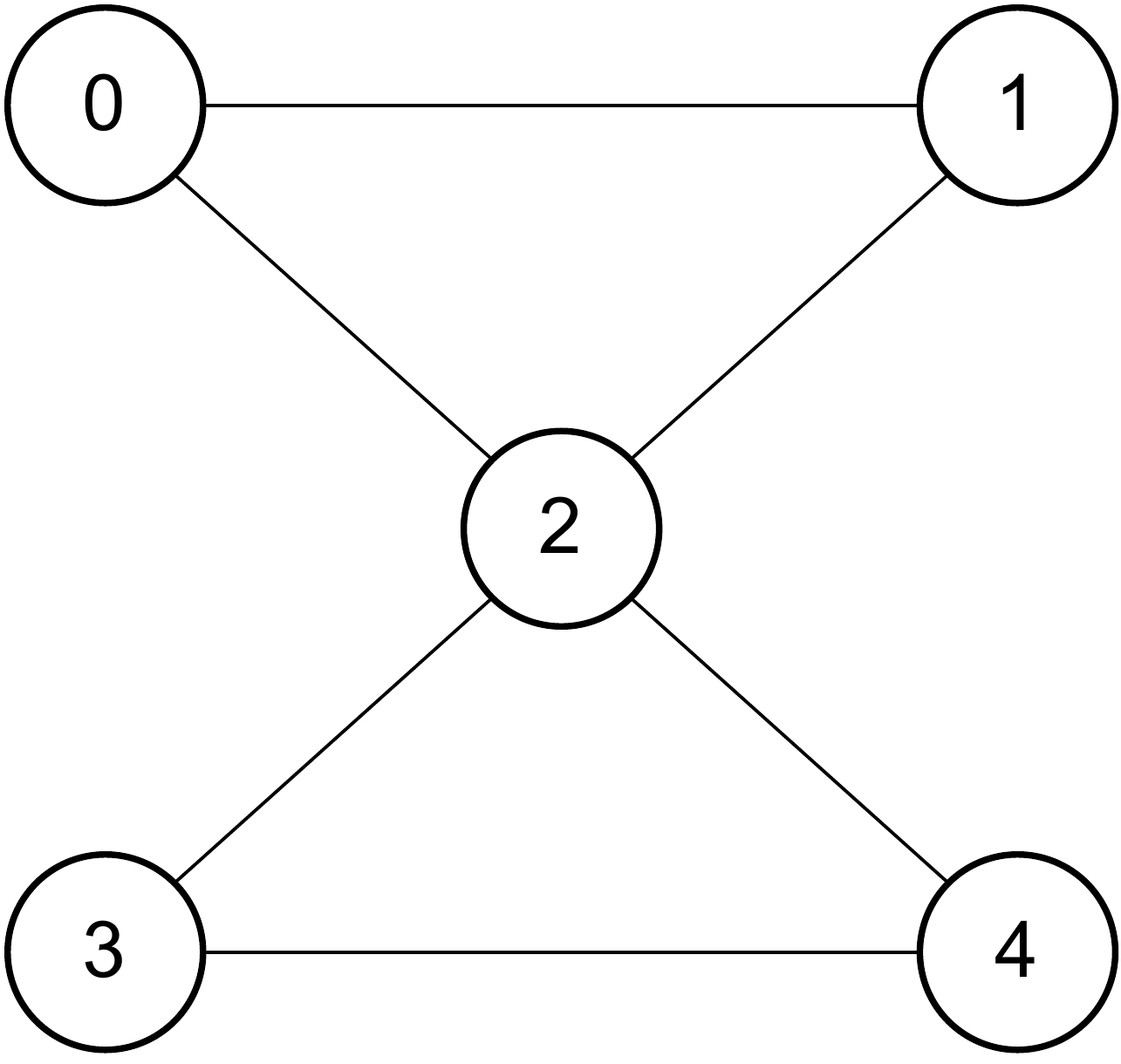}
\caption{Sketch of the 5-qubit simulated topology. All qubits are coupled for simplicity since they can all be concurrently controlled or targeted by controlled quantum operations.}
\label{fig:05}
\end{figure}

Similar to how we completed the noise-free simulation in Section II, we needed to import the logic gates we would use, as seen in Box 9.
\begin{tcolorbox}[breakable, size=fbox, boxrule=1pt, pad at break*=1mm,colback=cellbackground, colframe=cellborder,coltitle=black,title=Box 9: Importing the computational tools]
\begin{Verbatim}[commandchars=\\\{\}]
\PY{k+kn}{from} \PY{n+nn}{qat}\PY{n+nn}{.}\PY{n+nn}{lang}\PY{n+nn}{.}\PY{n+nn}{AQASM} \PY{k+kn}{import} \PY{n}{Program}\PY{p}{,} \PY{n}{H}\PY{p}{,} \PY{n}{X}\PY{p}{,} \PY{n}{CCNOT}
\end{Verbatim}
\end{tcolorbox}
In this scenario, the allocation of classical bits and quantum bits takes place in the exact same manner as it did in the previous simulation. The fact that we employ one of these qubits as an ancilla necessitates that we now assign 4 classical bits and 5 quantum bits in our system. (see Box 10).
\begin{tcolorbox}[breakable, size=fbox, boxrule=1pt, pad at break*=1mm,colback=cellbackground, colframe=cellborder,coltitle=black,title=Box 10: Allocating qubits and classical bits registers]
\begin{Verbatim}[commandchars=\\\{\}]
\PY{c+c1}{\PYZsh{} Creating Grover\PYZsq{}s program}
\PY{n}{grover} \PY{o}{=} \PY{n}{Program}\PY{p}{(}\PY{p}{)}

\PY{c+c1}{\PYZsh{} Allocating qubits and classical bits registers}
\PY{n}{Qr} \PY{o}{=} \PY{n}{grover}\PY{o}{.}\PY{n}{qalloc}\PY{p}{(}\PY{n}{5}\PY{p}{)}
\PY{n}{Cr} \PY{o}{=} \PY{n}{grover}\PY{o}{.}\PY{n}{calloc}\PY{p}{(}\PY{n}{4}\PY{p}{)}
\end{Verbatim}
\end{tcolorbox}

In order to build the topology connections for the quantum hardware (as shown in Fig. \ref{fig:05}), we must import the libraries \texttt{Topology} and \texttt{HardwareSpecs}. Box 11 contains the coding for the list items that represent the connections between the qubits in the idealized topology. The first member of each ordered pair is the control qubit, and the second element is the target one.

\begin{tcolorbox}[breakable, size=fbox, boxrule=1pt, pad at break*=1mm,colback=cellbackground, colframe=cellborder,coltitle=black,title=Box 11: Defining the topology of the simulated quantum hardware]
\begin{Verbatim}[commandchars=\\\{\}]
\PY{k+kn}{from} \PY{n+nn}{qat}\PY{n+nn}{.}\PY{n+nn}{core}\PY{n+nn}{.}\PY{n+nn}{wrappers}\PY{n+nn}{.}\PY{n+nn}{hardware\PYZus{}specs} \PY{k+kn}{import} \PY{n}{Topology}\PY{p}{,} \PY{n}{HardwareSpecs}

\PY{n}{topology} \PY{o}{=} \PY{n}{Topology}\PY{p}{(}\PY{p}{)}
\PY{k}{for} \PY{n}{control}\PY{p}{,} \PY{n}{target} \PY{o+ow}{in} \PY{p}{[}\PY{p}{(}\PY{l+m+mi}{0}\PY{p}{,} \PY{l+m+mi}{1}\PY{p}{)}\PY{p}{,} \PY{p}{(}\PY{l+m+mi}{0}\PY{p}{,} \PY{l+m+mi}{2}\PY{p}{)}\PY{p}{,} \PY{p}{(}\PY{l+m+mi}{1}\PY{p}{,} \PY{l+m+mi}{0}\PY{p}{)}\PY{p}{,} \PY{p}{(}\PY{l+m+mi}{1}\PY{p}{,} \PY{l+m+mi}{2}\PY{p}{)}\PY{p}{,}
                        \PY{p}{(}\PY{l+m+mi}{2}\PY{p}{,} \PY{l+m+mi}{0}\PY{p}{)}\PY{p}{,} \PY{p}{(}\PY{l+m+mi}{2}\PY{p}{,} \PY{l+m+mi}{1}\PY{p}{)}\PY{p}{,} \PY{p}{(}\PY{l+m+mi}{2}\PY{p}{,} \PY{l+m+mi}{3}\PY{p}{)}\PY{p}{,} \PY{p}{(}\PY{l+m+mi}{2}\PY{p}{,} \PY{l+m+mi}{4}\PY{p}{)}\PY{p}{,} 
                        \PY{p}{(}\PY{l+m+mi}{3}\PY{p}{,} \PY{l+m+mi}{2}\PY{p}{)}\PY{p}{,} \PY{p}{(}\PY{l+m+mi}{3}\PY{p}{,} \PY{l+m+mi}{4}\PY{p}{)}\PY{p}{,} \PY{p}{(}\PY{l+m+mi}{4}\PY{p}{,} \PY{l+m+mi}{2}\PY{p}{)}\PY{p}{,} \PY{p}{(}\PY{l+m+mi}{4}\PY{p}{,} \PY{l+m+mi}{3}\PY{p}{)}
                       \PY{p}{]}\PY{p}{:}
    \PY{n}{topology}\PY{o}{.}\PY{n}{add\PYZus{}edge}\PY{p}{(}\PY{n}{control}\PY{p}{,} \PY{n}{target}\PY{p}{)}
    
\PY{n}{hw\PYZus{}specs} \PY{o}{=} \PY{n}{HardwareSpecs}\PY{p}{(}\PY{n}{nbqbits}\PY{o}{=}\PY{n}{n}\PY{p}{,} \PY{n}{topology}\PY{o}{=}\PY{n}{topology}\PY{p}{)}
\end{Verbatim}
\end{tcolorbox}

In Box 12, the initialization of the qubits occurs during the first step of Grover's algorithm by putting them in a balanced superposition. Since qubit 2 is a supplementary qubit in this instance, it is unnecessary to start it in a superposition state.

\begin{tcolorbox}[breakable, size=fbox, boxrule=1pt, pad at break*=1mm,colback=cellbackground, colframe=cellborder,coltitle=black,title=Box 12: Initializing the qubits in a balanced superposition]
\begin{Verbatim}[commandchars=\\\{\}]
\PY{n}{H}\PY{p}{(}\PY{n}{Qr}\PY{p}{[}\PY{l+m+mi}{0}\PY{p}{]}\PY{p}{)}
\PY{n}{H}\PY{p}{(}\PY{n}{Qr}\PY{p}{[}\PY{l+m+mi}{1}\PY{p}{]}\PY{p}{)}
\PY{n}{H}\PY{p}{(}\PY{n}{Qr}\PY{p}{[}\PY{l+m+mi}{3}\PY{p}{]}\PY{p}{)}
\PY{n}{H}\PY{p}{(}\PY{n}{Qr}\PY{p}{[}\PY{l+m+mi}{4}\PY{p}{]}\PY{p}{)}
\end{Verbatim}
\end{tcolorbox}

The second and third phases of the quantum search algorithm, implemented on the simulated quantum hardware are shown in Box 13. Initially, we apply the oracle that explicitly looks for the state  $\ket{1111}$ by applying the MCZ gate decomposed as illustrated in Fig. \ref{fig:04} (c). Next, we use amplitude amplification to enhance the probability that the marked object will be discovered while decreasing the probability that other items. These two procedures are repeated $\sqrt{N}$ times, in order to maximize the probability of finding the sought state, in the same way as performed in the noise-free simulation in section II.

\begin{tcolorbox}[breakable, size=fbox, boxrule=1pt, pad at break*=1mm,colback=cellbackground, colframe=cellborder,coltitle=black,title=Box 13: Applying the Oracle and Amplitude Amplification subroutine that marks state $\ket{1111}$]
\begin{Verbatim}[commandchars=\\\{\}]
\PY{k}{for} \PY{n}{i} \PY{o+ow}{in} \PY{n+nb}{range}\PY{p}{(}\PY{l+m+mi}{3}\PY{p}{)}\PY{p}{:}
    \PY{c+c1}{\PYZsh{} Oracle}
    \PY{n}{H}\PY{p}{(}\PY{n}{Qr}\PY{p}{[}\PY{l+m+mi}{4}\PY{p}{]}\PY{p}{)}
    \PY{n}{CCNOT}\PY{p}{(}\PY{n}{Qr}\PY{p}{[}\PY{l+m+mi}{0}\PY{p}{]}\PY{p}{,} \PY{n}{Qr}\PY{p}{[}\PY{l+m+mi}{1}\PY{p}{]}\PY{p}{,} \PY{n}{Qr}\PY{p}{[}\PY{l+m+mi}{2}\PY{p}{]}\PY{p}{)}
    \PY{n}{CCNOT}\PY{p}{(}\PY{n}{Qr}\PY{p}{[}\PY{l+m+mi}{2}\PY{p}{]}\PY{p}{,} \PY{n}{Qr}\PY{p}{[}\PY{l+m+mi}{3}\PY{p}{]}\PY{p}{,} \PY{n}{Qr}\PY{p}{[}\PY{l+m+mi}{4}\PY{p}{]}\PY{p}{)}
    \PY{n}{CCNOT}\PY{p}{(}\PY{n}{Qr}\PY{p}{[}\PY{l+m+mi}{0}\PY{p}{]}\PY{p}{,} \PY{n}{Qr}\PY{p}{[}\PY{l+m+mi}{1}\PY{p}{]}\PY{p}{,} \PY{n}{Qr}\PY{p}{[}\PY{l+m+mi}{2}\PY{p}{]}\PY{p}{)}
    \PY{n}{H}\PY{p}{(}\PY{n}{Qr}\PY{p}{[}\PY{l+m+mi}{4}\PY{p}{]}\PY{p}{)}

    \PY{c+c1}{\PYZsh{} Amplitude Amplification}
    \PY{n}{H}\PY{p}{(}\PY{n}{Qr}\PY{p}{[}\PY{l+m+mi}{0}\PY{p}{]}\PY{p}{)}
    \PY{n}{H}\PY{p}{(}\PY{n}{Qr}\PY{p}{[}\PY{l+m+mi}{1}\PY{p}{]}\PY{p}{)}
    \PY{n}{H}\PY{p}{(}\PY{n}{Qr}\PY{p}{[}\PY{l+m+mi}{3}\PY{p}{]}\PY{p}{)}
    \PY{n}{H}\PY{p}{(}\PY{n}{Qr}\PY{p}{[}\PY{l+m+mi}{4}\PY{p}{]}\PY{p}{)}
    \PY{n}{X}\PY{p}{(}\PY{n}{Qr}\PY{p}{[}\PY{l+m+mi}{0}\PY{p}{]}\PY{p}{)}
    \PY{n}{X}\PY{p}{(}\PY{n}{Qr}\PY{p}{[}\PY{l+m+mi}{1}\PY{p}{]}\PY{p}{)}
    \PY{n}{X}\PY{p}{(}\PY{n}{Qr}\PY{p}{[}\PY{l+m+mi}{3}\PY{p}{]}\PY{p}{)}
    \PY{n}{X}\PY{p}{(}\PY{n}{Qr}\PY{p}{[}\PY{l+m+mi}{4}\PY{p}{]}\PY{p}{)}
    \PY{n}{H}\PY{p}{(}\PY{n}{Qr}\PY{p}{[}\PY{l+m+mi}{4}\PY{p}{]}\PY{p}{)}
    \PY{n}{CCNOT}\PY{p}{(}\PY{n}{Qr}\PY{p}{[}\PY{l+m+mi}{0}\PY{p}{]}\PY{p}{,} \PY{n}{Qr}\PY{p}{[}\PY{l+m+mi}{1}\PY{p}{]}\PY{p}{,} \PY{n}{Qr}\PY{p}{[}\PY{l+m+mi}{2}\PY{p}{]}\PY{p}{)}
    \PY{n}{CCNOT}\PY{p}{(}\PY{n}{Qr}\PY{p}{[}\PY{l+m+mi}{2}\PY{p}{]}\PY{p}{,} \PY{n}{Qr}\PY{p}{[}\PY{l+m+mi}{3}\PY{p}{]}\PY{p}{,} \PY{n}{Qr}\PY{p}{[}\PY{l+m+mi}{4}\PY{p}{]}\PY{p}{)}
    \PY{n}{CCNOT}\PY{p}{(}\PY{n}{Qr}\PY{p}{[}\PY{l+m+mi}{0}\PY{p}{]}\PY{p}{,} \PY{n}{Qr}\PY{p}{[}\PY{l+m+mi}{1}\PY{p}{]}\PY{p}{,} \PY{n}{Qr}\PY{p}{[}\PY{l+m+mi}{2}\PY{p}{]}\PY{p}{)}
    \PY{n}{H}\PY{p}{(}\PY{n}{Qr}\PY{p}{[}\PY{l+m+mi}{4}\PY{p}{]}\PY{p}{)}
    \PY{n}{X}\PY{p}{(}\PY{n}{Qr}\PY{p}{[}\PY{l+m+mi}{0}\PY{p}{]}\PY{p}{)}
    \PY{n}{X}\PY{p}{(}\PY{n}{Qr}\PY{p}{[}\PY{l+m+mi}{1}\PY{p}{]}\PY{p}{)}
    \PY{n}{X}\PY{p}{(}\PY{n}{Qr}\PY{p}{[}\PY{l+m+mi}{3}\PY{p}{]}\PY{p}{)}
    \PY{n}{X}\PY{p}{(}\PY{n}{Qr}\PY{p}{[}\PY{l+m+mi}{4}\PY{p}{]}\PY{p}{)}
    \PY{n}{H}\PY{p}{(}\PY{n}{Qr}\PY{p}{[}\PY{l+m+mi}{0}\PY{p}{]}\PY{p}{)}
    \PY{n}{H}\PY{p}{(}\PY{n}{Qr}\PY{p}{[}\PY{l+m+mi}{1}\PY{p}{]}\PY{p}{)}
    \PY{n}{H}\PY{p}{(}\PY{n}{Qr}\PY{p}{[}\PY{l+m+mi}{3}\PY{p}{]}\PY{p}{)}
    \PY{n}{H}\PY{p}{(}\PY{n}{Qr}\PY{p}{[}\PY{l+m+mi}{4}\PY{p}{]}\PY{p}{)}
\end{Verbatim}
\end{tcolorbox}

After this, we will finish the last step of Grover's algorithm, which consists of taking measurements in accordance with Box 14. It is worth noting that, the execution of the measurement on the auxiliary qubit is completely unnecessary.

\begin{tcolorbox}[breakable, size=fbox, boxrule=1pt, pad at break*=1mm,colback=cellbackground, colframe=cellborder,coltitle=black,title=Box 14: Measurements]
\begin{Verbatim}[commandchars=\\\{\}]
\PY{n}{grover}\PY{o}{.}\PY{n}{measure}\PY{p}{(}\PY{n}{Qr}\PY{p}{[}\PY{l+m+mi}{0}\PY{p}{]}\PY{p}{,} \PY{n}{Cr}\PY{p}{[}\PY{l+m+mi}{0}\PY{p}{]}\PY{p}{)}
\PY{n}{grover}\PY{o}{.}\PY{n}{measure}\PY{p}{(}\PY{n}{Qr}\PY{p}{[}\PY{l+m+mi}{1}\PY{p}{]}\PY{p}{,} \PY{n}{Cr}\PY{p}{[}\PY{l+m+mi}{1}\PY{p}{]}\PY{p}{)}
\PY{n}{grover}\PY{o}{.}\PY{n}{measure}\PY{p}{(}\PY{n}{Qr}\PY{p}{[}\PY{l+m+mi}{3}\PY{p}{]}\PY{p}{,} \PY{n}{Cr}\PY{p}{[}\PY{l+m+mi}{2}\PY{p}{]}\PY{p}{)}
\PY{n}{grover}\PY{o}{.}\PY{n}{measure}\PY{p}{(}\PY{n}{Qr}\PY{p}{[}\PY{l+m+mi}{4}\PY{p}{]}\PY{p}{,} \PY{n}{Cr}\PY{p}{[}\PY{l+m+mi}{3}\PY{p}{]}\PY{p}{)}

\PY{c+c1}{\PYZsh{} Returning a circuit implementing Grover\PYZsq{}s program}
\PY{n}{circuit} \PY{o}{=} \PY{n}{grover}\PY{o}{.}\PY{n}{to\PYZus{}circ}\PY{p}{(}\PY{p}{)}

\PY{c+c1}{\PYZsh{} Drawing the circuit}
\PY{o}{\PYZpc{}}\PY{k}{qatdisplay} circuit  \PYZhy{}\PYZhy{}svg
\end{Verbatim}
\end{tcolorbox}

\subsection{Quantum Noise}

In the following, we introduce the quantum noise model in the simulated quantum hardware. Generally, qubit relaxation time $T_{1}$ (i.e. \textit{longitudinal relaxation} or \textit{amplitude damping}) and qubit dephasing time $T_{2}$ (i.e. \textit{transverse relaxation}) are used to quantify quantum noise in quantum hardware setups\cite{youssef2020measuring, rost2020simulation,L__2007}. On the other hand, the Pure Dephasing time 
\begin{equation}
T_{\phi} = \frac{1}{\frac{1}{T_2} - \frac{1}{2T_1}}    
\end{equation}
is often the dominant contribution to $T_{2}$ \cite{skinner1986pure,ferraro2019phonon}. This parameter determines how long a system will be able to keep its coherence, whereas amplitude damping offers a model for the physical process of energy decay associated with the application of quantum noise. \cite{Dudley:10}. %We have applied the Pure Dephasing channel, where \cite{}:%We consider only the existence of pure dephasing in idle time, not introducing noise inserted by the application of the gates and measurements. Parameter dephasing determines how long a system will maintain its coherence, while amplitude damping provides a model for the physical process of energy decay associated with the application of noise \cite{Dudley:10}. The parameter pure dephasing is often the dominant contribution to 1/T2 \cite{skinner1986pure}. We have applied the Pure Dephasing channel, where \cite{ferraro2019phonon}
In this regard, we introduce the Pure Dephasing channel in order to perform the noisy simulations on the quantum processor emulated with the topology described in Fig. \ref{fig:05}. 
Furthermore, the operating time for the quantum gates used in building quantum hardware topology must be specified. Box 15 imports the library \texttt{DefaultGatesSpecification} and specifies the application timings of the gates used in Grover's algorithm. Therefore, in the total amount of time required to complete the procedures, we take into consideration the following factors: \texttt{X} gate $\rightarrow$ 35.5 ns; \texttt{Hadamard} gate $\rightarrow$ 35.5 ns;  \texttt{CCNOT} gate $\rightarrow$ 350 ns, Measurements $\rightarrow$ 35.5 ns.

%\textcolor{red}{At this point, we need to define the application time of the gates, as well as the time to perform a measurement on the qubits. In Box 15 we import the ``DefaultGatesSpecification" library and define the application times of the gates as their specifications.}

% \begin{tcolorbox}[breakable, size=fbox, boxrule=1pt, pad at break*=1mm,colback=cellbackground, colframe=cellborder,coltitle=black,title=Box 15: Simulating and plotting the results]
% \begin{Verbatim}[commandchars=\\\{\}]
% \PY{c+c1}{\PYZsh{} Simulating and plotting the results}
% \PY{k+kn}{from} \PY{n+nn}{qat}\PY{n+nn}{.}\PY{n+nn}{qpus} \PY{k+kn}{import} \PY{n}{PyLinalg}
% \PY{n}{qpu}\PY{o}{=}\PY{n}{PyLinalg}\PY{p}{(}\PY{p}{)}
% \PY{n}{job} \PY{o}{=} \PY{n}{circuit}\PY{o}{.}\PY{n}{to\PYZus{}job}\PY{p}{(}\PY{n}{nbshots}\PY{o}{=}\PY{l+m+mi}{1024}\PY{p}{)}
% \PY{n}{result} \PY{o}{=} \PY{n}{qpu}\PY{o}{.}\PY{n}{submit}\PY{p}{(}\PY{n}{job}\PY{p}{)}

% \PY{c+c1}{\PYZsh{} Printing the results}
% \PY{k}{for} \PY{n}{sample} \PY{o+ow}{in} \PY{n}{result}\PY{p}{:}
%     \PY{n+nb}{print}\PY{p}{(}\PY{l+s+s2}{\PYZdq{}}\PY{l+s+s2}{State }\PY{l+s+si}{\PYZpc{}s}\PY{l+s+s2}{: probability }\PY{l+s+si}{\PYZpc{}s}\PY{l+s+s2}{ +/\PYZhy{} }\PY{l+s+si}{\PYZpc{}s}\PY{l+s+s2}{\PYZdq{}} \PY{o}{\PYZpc{}} \PY{p}{(}\PY{n}{sample}\PY{o}{.}\PY{n}{state}\PY{p}{,} \PY{n}{sample}\PY{o}{.}\PY{n}{probability}\PY{p}{,} \PY{n}{sample}\PY{o}{.}\PY{n}{err}\PY{p}{)}\PY{p}{)}
% \end{Verbatim}
% \end{tcolorbox}

\begin{tcolorbox}[breakable, size=fbox, boxrule=1pt, pad at break*=1mm,colback=cellbackground, colframe=cellborder,coltitle=black,title=Box 15: Defining gate times]
\begin{Verbatim}[commandchars=\\\{\}]
\PY{n}{gate\PYZus{}times} \PY{o}{=} \PY{p}{\PYZob{}}\PY{l+s+s2}{\PYZdq{}}\PY{l+s+s2}{X}\PY{l+s+s2}{\PYZdq{}}\PY{p}{:}\PY{l+m+mf}{35.5}\PY{p}{,} \PY{l+s+s2}{\PYZdq{}}\PY{l+s+s2}{H}\PY{l+s+s2}{\PYZdq{}}\PY{p}{:}\PY{l+m+mf}{35.5}\PY{p}{,} \PY{l+s+s2}{\PYZdq{}}\PY{l+s+s2}{CCNOT}\PY{l+s+s2}{\PYZdq{}}\PY{p}{:}\PY{l+m+mi}{350}\PY{p}{,} \PY{l+s+s2}{\PYZdq{}}\PY{l+s+s2}{measure}\PY{l+s+s2}{\PYZdq{}}\PY{p}{:}\PY{l+m+mf}{35.5}\PY{p}{\PYZcb{}} 

\PY{k+kn}{from} \PY{n+nn}{qat}\PY{n+nn}{.}\PY{n+nn}{hardware} \PY{k+kn}{import} \PY{n}{DefaultGatesSpecification}
\PY{n}{gates\PYZus{}spec} \PY{o}{=} \PY{n}{DefaultGatesSpecification}\PY{p}{(}\PY{n}{gate\PYZus{}times}\PY{p}{)}
\end{Verbatim}
\end{tcolorbox}

Furthermore, in order to apply the Pure Dephasing channel in Box 16, we need to import the library known as \texttt{ParametricPureDephasing}. In addition to the noise model, one can define the qubit relaxation times $T_1$ and $T_2$, respectively as shown in Box 16.

\begin{tcolorbox}[breakable, size=fbox, boxrule=1pt, pad at break*=1mm,colback=cellbackground, colframe=cellborder,coltitle=black,title=Box 16: Defining qubit relaxation times]
\begin{Verbatim}[commandchars=\\\{\}]
\PY{k+kn}{from} \PY{n+nn}{qat}\PY{n+nn}{.}\PY{n+nn}{quops} \PY{k+kn}{import} \PY{n}{ParametricPureDephasing}

\PY{n}{T1}\PY{p}{,} \PY{n}{T2} \PY{o}{=} \PY{l+m+mi}{1000}\PY{p}{,} \PY{l+m+mi}{1000} \PY{c+c1}{\PYZsh{} nanosecs}

\PY{c+c1}{\PYZsh{} Pure Dephasing: Lindblad approximation.}
\PY{n}{PD\PYZus{}Lindblad} \PY{o}{=} \PY{n}{ParametricPureDephasing}\PY{p}{(}\PY{n}{T\PYZus{}phi} \PY{o}{=} \PY{l+m+mi}{1}\PY{o}{/}\PY{p}{(}\PY{l+m+mi}{1}\PY{o}{/}\PY{n}{T2} \PY{o}{\PYZhy{}} \PY{l+m+mi}{1}\PY{o}{/}\PY{p}{(}\PY{l+m+mi}{2}\PY{o}{*}\PY{n}{T1}\PY{p}{)}\PY{p}{)}\PY{p}{)}
\end{Verbatim}
\end{tcolorbox}

Subsequently, after importing the \texttt{HardwareModel} and \texttt{NoisyQProc} libraries, we specify the model of the quantum processor that we will use, including the gate specifications and the noise model that will be applied (see Box 17).

\begin{tcolorbox}[breakable, size=fbox, boxrule=1pt, pad at break*=1mm,colback=cellbackground, colframe=cellborder,coltitle=black,title=Box 17: Importing quantum hardware]
\begin{Verbatim}[commandchars=\\\{\}]
\PY{k+kn}{from} \PY{n+nn}{qat}\PY{n+nn}{.}\PY{n+nn}{hardware} \PY{k+kn}{import} \PY{n}{HardwareModel}
\PY{k+kn}{from} \PY{n+nn}{qat}\PY{n+nn}{.}\PY{n+nn}{qpus} \PY{k+kn}{import} \PY{n}{NoisyQProc}

\PY{n}{hardware\PYZus{}lindblad} \PY{o}{=} \PY{n}{HardwareModel}\PY{p}{(}\PY{n}{gates\PYZus{}spec}\PY{p}{,} \PY{n}{idle\PYZus{}noise}\PY{o}{=}\PY{p}{[}\PY{n}{PD\PYZus{}Lindblad}\PY{p}{]}\PY{p}{)}
\PY{n}{noisy\PYZus{}qpu} \PY{o}{=} \PY{n}{NoisyQProc}\PY{p}{(}\PY{n}{hardware\PYZus{}model}\PY{o}{=}\PY{n}{hardware\PYZus{}lindblad}\PY{p}{)}
\end{Verbatim}
\end{tcolorbox}

Finally, Grover's algorithm in the simulated quantum hardware under the Pure Dephasing noise model, as presented in Box 18.
\begin{tcolorbox}[breakable, size=fbox, boxrule=1pt, pad at break*=1mm,colback=cellbackground, colframe=cellborder,coltitle=black,title=Box 18: Simulation]
\begin{Verbatim}[commandchars=\\\{\}]
\PY{n}{job} \PY{o}{=} \PY{n}{circuit}\PY{o}{.}\PY{n}{to\PYZus{}job}\PY{p}{(}\PY{n}{nbshots}\PY{o}{=}\PY{l+m+mi}{8192}, qubits=[0,1,3,4]\PY{p}{)}
\PY{n}{result} \PY{o}{=} \PY{n}{noisy\PYZus{}qpu}\PY{o}{.}\PY{n}{submit}\PY{p}{(}\PY{n}{job}\PY{p}{)}
\PY{k}{for} \PY{n}{sample} \PY{o+ow}{in} \PY{n}{result}\PY{p}{:}
    \PY{n+nb}{print}\PY{p}{(}\PY{l+s+s2}{\PYZdq{}}\PY{l+s+s2}{State }\PY{l+s+si}{\PYZpc{}s}\PY{l+s+s2}{, probability }\PY{l+s+si}{\PYZpc{}s}\PY{l+s+s2}{, err }\PY{l+s+si}{\PYZpc{}s}\PY{l+s+s2}{\PYZdq{}}\PY{o}{\PYZpc{}}\PY{p}{(}\PY{n}{sample}\PY{o}{.}\PY{n}{state}\PY{p}{,} \PY{n}{sample}\PY{o}{.}\PY{n}{probability}\PY{p}{,} \PY{n}{sample}\PY{o}{.}\PY{n}{err}\PY{p}{)}\PY{p}{)}
\end{Verbatim}
\end{tcolorbox}

The gate application and relaxation times are properties of each quantum computer, which can be altered once the calibrations are performed. Therefore, we selected these settings not with the objective of recreating the results of a particular system but rather with the only purpose of demonstrating the impact of these noises on the outcomes of the simulation. 

The simulations were conducted out using the stochastic technique, which requires that the density matrix be interpreted using a probability distribution based on pure states \cite{nielsen2002quantum}. 
Using this method, the error scales with $\frac{1}{\sqrt{shots}}$. Thus, we perform the simulation with $2^{13}$ shots to get proper results.

Fig.~\ref{fig:06} shows the probability distributions of Grover's algorithm performed in quantum hardware simulated under the Pure Dephasing noise model. As can be seen, we can gradually observe the noise effects on the application of the algorithm by decreasing the  dephasing time ($T_{2}$). It is worth noting that, although the most probable state is the searched item $\ket{1111}$, the other states 
 arose with significant probabilities even after the $\sqrt{N}$ repetitions of the Oracle and Amplitude Amplification subroutines,  different from the noise-free simulation presented in Fig. \ref{fig:03}. As expected, the situation becomes worse when we decrease the dephasing time. The effect of the noise is to reduce the algorithm's accuracy, approximating the probability of the sought state to the other items of the unstructured list. 

\begin{figure}[H]
    \centering
    \subfigure[]{\includegraphics[scale=0.3]{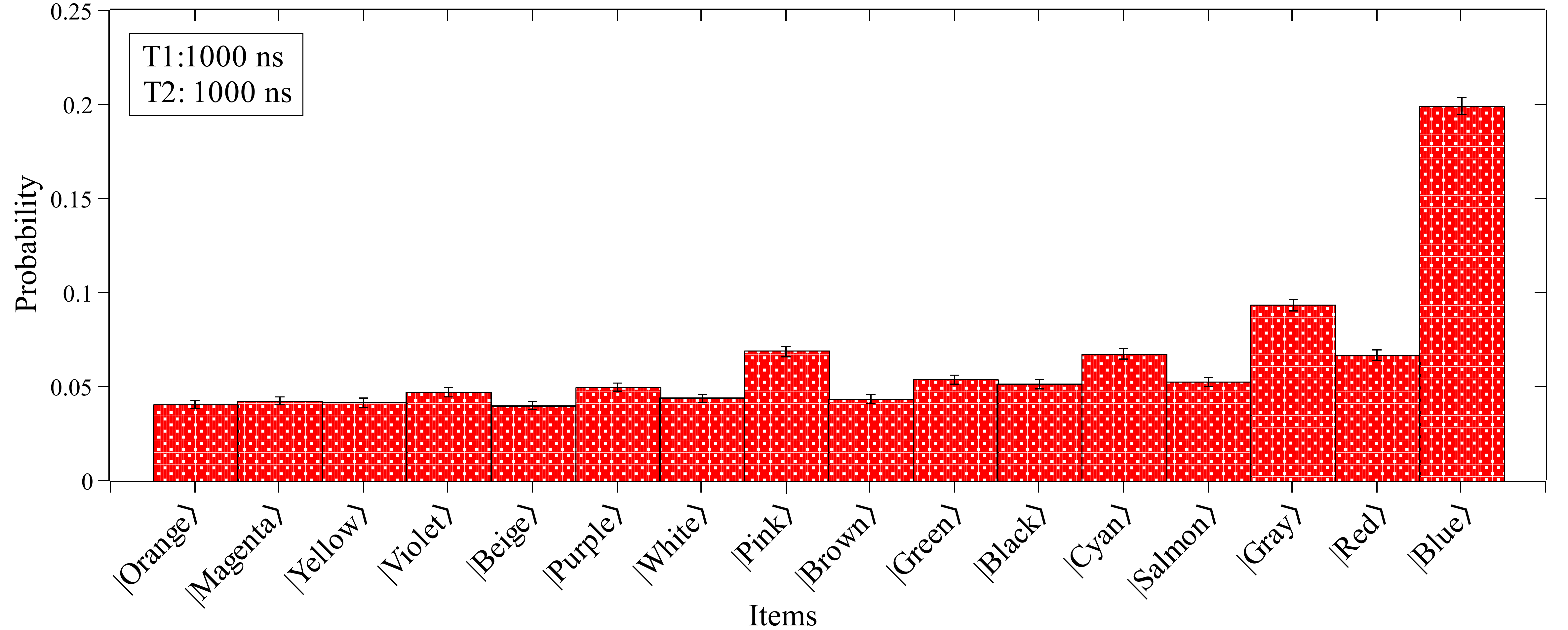}}\\
    \subfigure[]{\includegraphics[scale=0.3]{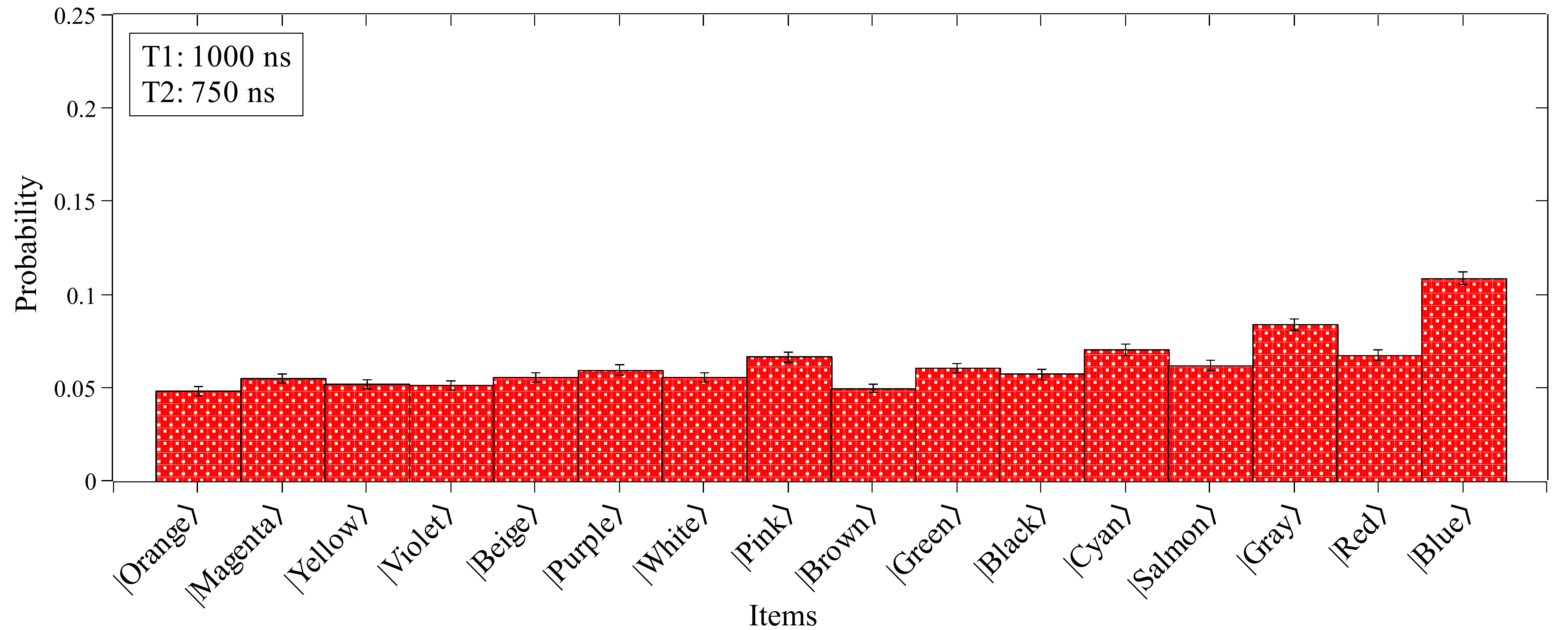}}
    \subfigure[]{\includegraphics[scale=0.3]{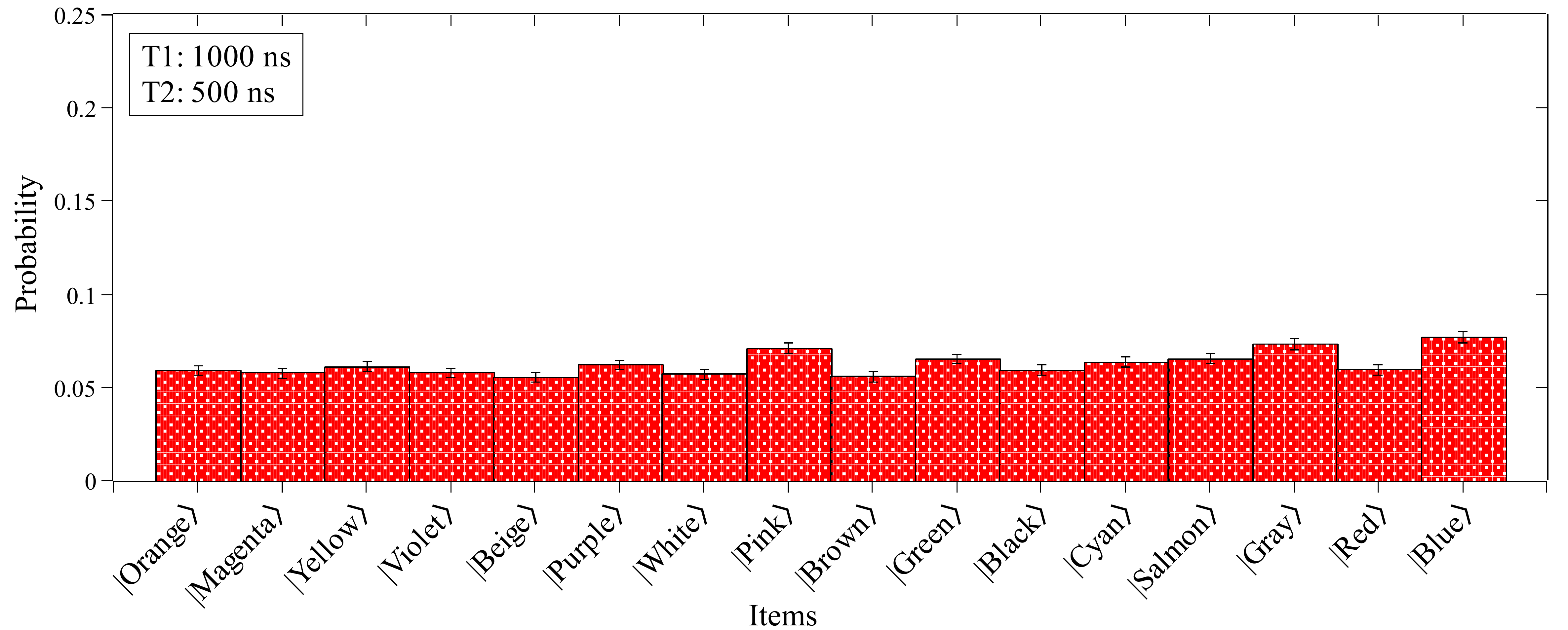}}
    \caption{Simulation with noise using $T_{phi}$ with  $T_1$ equal to 1000 ns and $T_2$ equal to (a) 1000 ns, (b) 750 ns, and (c) 500 ns.}
    \label{fig:06}
\end{figure}

In order to obtain a landscape of the effect of the Pure Dephasing channel on the sought state, we plot the probability of the sought state as a function of the Pure Dephasing times ($T_{\phi}$). As can be seen, increasing the $T_{\phi}$ the probability of the sought state approximates to a probability limit 95.86\%, obtained for the noise-free simulation presented in Fig. \ref{fig:03}, highlighted in the dashed blue line. 
On the other hand, the performance of the search algorithm is negatively impacted by decreasing the dephasing time, rapidly decreasing the probability sought state. Therefore, we are convinced that these findings demonstrate that the existence of quantum noise in quantum processors reduces the efficiency of the algorithm, which can lead to inaccurate outcomes when simulating quantum circuits.

\begin{figure}[H]
\centering
\includegraphics[scale=0.6]{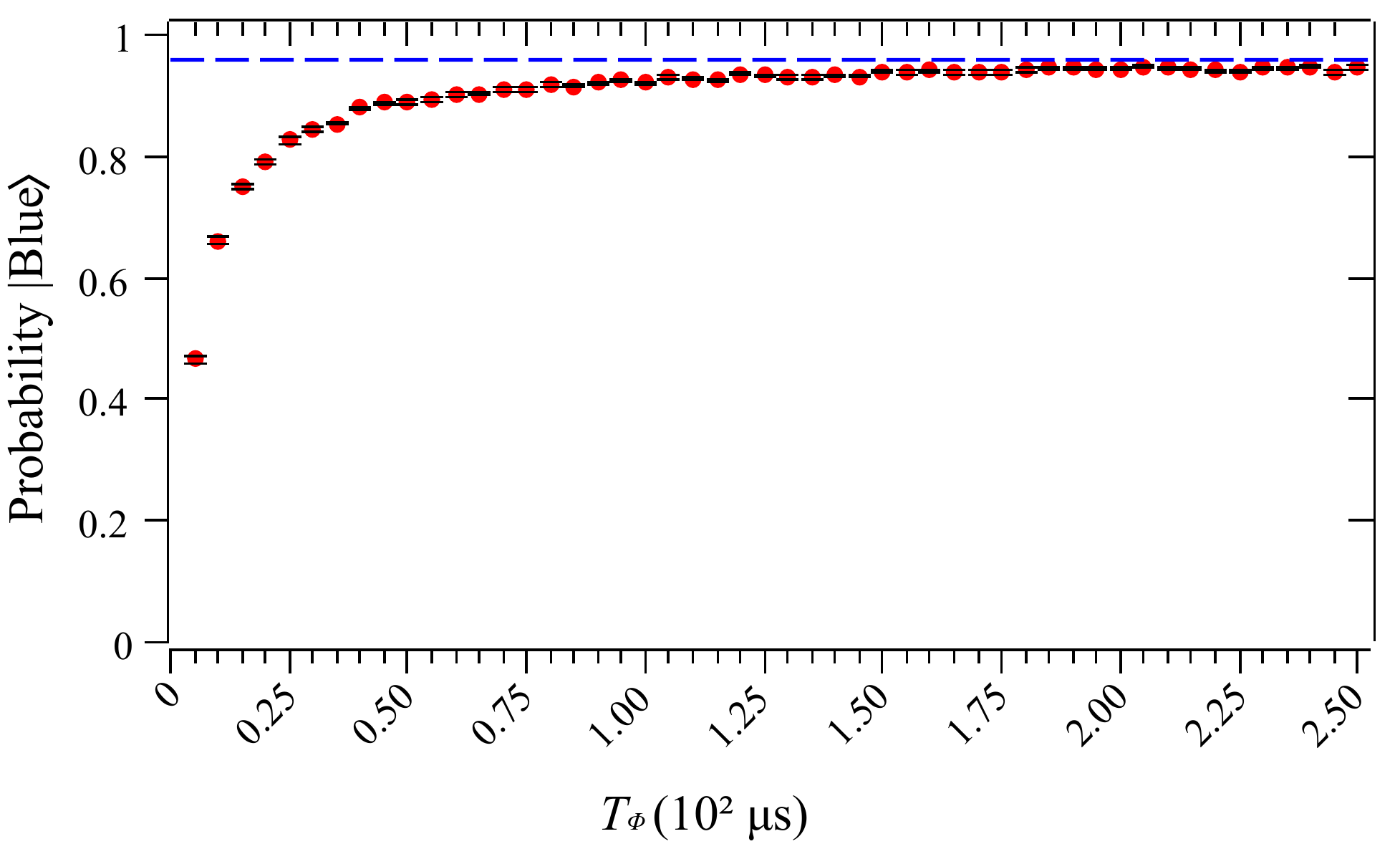}
\caption{Probability distribution for item $\ket{Blue}$ as a function of Pure Dephasing times, $T_{\phi}$. The dashed blue line highlights the probability of the sought state (95.86\%) obtained in the noise-free simulation presented in Fig. \ref{fig:03}.}
\label{fig:07}
\end{figure}

\section{Conclusion}
This paper presents the quantum search problem based on the famous Grover's algorithm associated with the binary encoding of words into quantum states of the 4-qubits computational basis.  We highlight the main conditions for developing projects and their implementation in dedicated hardware processors, showing how this can serve as an effective tool for teaching quantum computing in a practical way, allowing students to become active agents involved in the construction of knowledge. Our findings are consistent with the theoretical predictions found in the aforementioned literature for the examples that were covered, and they demonstrate that AQASM and myQLM  would be practical tools for both the implementation and analysis of quantum algorithms, as well as for the simulation of quantum hardware topologies, acting as a high-level quantum programming platform. Moreover, we emulate a genuine quantum processor by configuring a dedicated quantum simulator to conform to the necessary architectural specifications for performing Grover's algorithm.  Furthermore, we demonstrate that the existence of quantum noise leads to a progressive decline in the quality of the outcomes produced by a quantum noisy simulation when compared to the result performed in a noise-free setup. Finally, we propose as a potential direction for future study the scaling up of our program to a greater number of qubits, which would result in an exponential increase in our database capacity. 

\section*{Acknowledgments}

The authors thank SENAI CIMATEC for the access to the KUATOMU quantum simulator, which was used in obtaining the quantum noise simulation results.


\section*{References}

\begin{thebibliography}{10}

\bibitem{maia2019experimento}
Angevaldo~Menezes Maia~Filho and Indianara Silva.
\newblock O experimento ws de 1950 e as suas implica{\c{c}}{\~o}es para a
  segunda revolu{\c{c}}{\~a}o da mec{\^a}nica qu{\^a}ntica.
\newblock {\em Revista Brasileira de Ensino de F{\'\i}sica}, 41(2), 2019.

\bibitem{terhal2018quantum}
Barbara~M Terhal.
\newblock Quantum supremacy, here we come.
\newblock {\em Nature Physics}, 14(6):530--531, 2018.

\bibitem{harrow2017quantum}
Aram~W Harrow and Ashley Montanaro.
\newblock Quantum computational supremacy.
\newblock {\em Nature}, 549(7671):203--209, 2017.

\bibitem{PRXQuantum.1.020101}
Ivan~H. Deutsch.
\newblock Harnessing the power of the second quantum revolution.
\newblock {\em PRX Quantum}, 1:020101, Nov 2020.

\bibitem{atzori2019second}
Matteo Atzori and Roberta Sessoli.
\newblock The second quantum revolution: role and challenges of molecular
  chemistry.
\newblock {\em Journal of the American Chemical Society}, 141(29):11339--11352,
  2019.

\bibitem{gibney2019quantum}
Elizabeth Gibney.
\newblock The quantum gold rush.
\newblock {\em Nature}, 574(7776):22--24, 2019.

\bibitem{peterssen2020quantum}
Guido Peterssen.
\newblock Quantum technology impact: The necessary workforce for developing
  quantum software.
\newblock In {\em QANSWER}, pages 6--22, 2020.

\bibitem{vietz2021decision}
Daniel Vietz, Johanna Barzen, Frank Leymann, and Karoline Wild.
\newblock On decision support for quantum application developers:
  categorization, comparison, and analysis of existing technologies.
\newblock In {\em International Conference on Computational Science}, pages
  127--141. Springer, 2021.

\bibitem{gyongyosi2019survey}
Laszlo Gyongyosi and Sandor Imre.
\newblock A survey on quantum computing technology.
\newblock {\em Computer Science Review}, 31:51--71, 2019.

\bibitem{raymer2019us}
Michael~G Raymer and Christopher Monroe.
\newblock The us national quantum initiative.
\newblock {\em Quantum Science and Technology}, 4(2):020504, 2019.

\bibitem{aiello2021achieving}
Clarice~D Aiello, DD~Awschalom, Hannes Bernien, Tina Brower, Kenneth~R Brown,
  Todd~A Brun, Justin~R Caram, Eric Chitambar, Rosa Di~Felice, Karina~Montilla
  Edmonds, et~al.
\newblock Achieving a quantum smart workforce.
\newblock {\em Quantum Science and Technology}, 6(3):030501, 2021.

\bibitem{riedel2019europe}
Max Riedel, Matyas Kovacs, Peter Zoller, J{\"u}rgen Mlynek, and Tommaso
  Calarco.
\newblock Europe’s quantum flagship initiative.
\newblock {\em Quantum Science and Technology}, 4(2):020501, 2019.

\bibitem{macquarrie2020emerging}
Evan~R MacQuarrie, Christoph Simon, Stephanie Simmons, and Elicia Maine.
\newblock The emerging commercial landscape of quantum computing.
\newblock {\em Nature Reviews Physics}, 2(11):596--598, 2020.

\bibitem{amin2019needs}
Mohammad~N Amin, Ronald~P Uhlig, Pradip~Peter Dey, Bhaskar Sinha, and Shatha
  Jawad.
\newblock The needs and challenges of workforce development in quantum
  computing.
\newblock In {\em 2019 Pacific Southwest Section Meeting}, 2019.

\bibitem{santos2017computador}
Alan~C Santos.
\newblock O computador qu{\^a}ntico da ibm e o ibm quantum experience.
\newblock {\em Revista Brasileira de Ensino de F{\'\i}sica}, 39(1), 2017.

\bibitem{rabelo2018abordagem}
Wilson~RM Rabelo and Maria L{\'u}cia~M Costa.
\newblock Uma abordagem pedag{\'o}gica no ensino da computa{\c{c}}{\~a}o
  qu{\^a}ntica com um processador qu{\^a}ntico de 5-qbits.
\newblock {\em Revista Brasileira de Ensino de F{\'\i}sica}, 40(4), 2018.

\bibitem{alves2020simulating}
{\'E}merson~M Alves, Francisco~DS Gomes, H{\'e}rcules~S Santana, and Alan~C
  Santos.
\newblock Simulating single-spin dynamics on an ibm five-qubit chip.
\newblock {\em Revista Brasileira de Ensino de F{\'\i}sica}, 42, 2020.

\bibitem{perry2019quantum}
Anastasia Perry, Ranbel Sun, Ciaran Hughes, Joshua Isaacson, and Jessica
  Turner.
\newblock Quantum computing as a high school module.
\newblock {\em arXiv preprint arXiv:1905.00282}, 2019.

\bibitem{tappert2019experience}
Charles~C Tappert, Ronald~I Frank, Istvan Barabasi, Avery~M Leider, Daniel
  Evans, and Lewis Westfall.
\newblock Experience teaching quantum computing.
\newblock In {\em 2019 ASCUE Proceedings}. Association Supporting Computer
  Users in Education, 2019.

\bibitem{jesus2021computaccao}
Gleydson Fernandes~de Jesus, Maria Helo{\'\i}sa~Fraga da~Silva,
  Teonas~Gon{\c{c}}alves Dourado~Netto, Lucas~Queiroz Galv{\~a}o,
  Frankle~Gabriel de~Oliveira~Souza, and Clebson Cruz.
\newblock Computa{\c{c}}{\~a}o qu{\^a}ntica: uma abordagem para a
  gradua{\c{c}}{\~a}o usando o qiskit.
\newblock {\em Revista Brasileira de Ensino de F{\'\i}sica}, 43, 2021.

\bibitem{candela2015undergraduate}
D~Candela.
\newblock Undergraduate computational physics projects on quantum computing.
\newblock {\em American Journal of Physics}, 83(8):688--702, 2015.

\bibitem{angara2020quantum}
Prashanti~Priya Angara, Ulrike Stege, and Andrew MacLean.
\newblock Quantum computing for high-school students an experience report.
\newblock In {\em 2020 IEEE International Conference on Quantum Computing and
  Engineering (QCE)}, pages 323--329. IEEE, 2020.

\bibitem{uhlig2019generating}
Ronald~P Uhlig, Pradip~Peter Dey, Shatha Jawad, Bhaskar~Raj Sinha, and Mohammad
  Amin.
\newblock Generating student interest in quantum computing.
\newblock In {\em 2019 IEEE Frontiers in Education Conference (FIE)}, pages
  1--9. IEEE, 2019.

\bibitem{billig2018quantum}
Yuly Billig.
\newblock Quantum computing for high school students.
\newblock {\em Yuly Billig, Ottawa}, 2018.

\bibitem{sutor2019dancing}
Robert~S Sutor.
\newblock {\em Dancing with Qubits: How quantum computing works and how it can
  change the world}.
\newblock Packt Publishing Ltd, 2019.

\bibitem{westfall2018teaching}
Lewis Westfall and Avery Leider.
\newblock Teaching quantum computing.
\newblock In {\em Proceedings of the Future Technologies Conference}, pages
  63--80. Springer, 2018.

\bibitem{faria}
E.~V. Faria.
\newblock A tecnologia da informação e da comunicação como ferramenta para
  a construção e democratização do conhecimento.
\newblock {\em Revista Scientia FAER}, 1(1):18, 2009.

\bibitem{wing2008computational}
Jeannette~M Wing.
\newblock Computational thinking and thinking about computing.
\newblock {\em Philosophical Transactions of the Royal Society A: Mathematical,
  Physical and Engineering Sciences}, 366(1881):3717--3725, 2008.

\bibitem{castillo2019classical}
Jairo~Ernesto Castillo, Yesenia Sierra, and Nelson~L Cubillos.
\newblock Classical simulation of grovers quantum algorithm.
\newblock {\em Revista Brasileira de Ensino de F{\'\i}sica}, 42, 2019.

\bibitem{leider2019quantum}
Avery Leider, Sadida Siddiqui, Daniel~A Sabol, and Charles~C Tappert.
\newblock Quantum computer search algorithms: Can we outperform the classical
  search algorithms?
\newblock In {\em Proceedings of the Future Technologies Conference}, pages
  447--459. Springer, 2019.

\bibitem{nielsen2002quantum}
Michael~A Nielsen and Isaac Chuang.
\newblock Quantum computation and quantum information, 2002.

\bibitem{castillo2020classical}
Jairo~Ernesto Castillo, Yesenia Sierra, and Nelson~L Cubillos.
\newblock Classical simulation of grovers quantum algorithm.
\newblock {\em Revista Brasileira de Ensino de F{\'\i}sica}, 42, 2020.

\bibitem{figgatt2017complete}
Caroline Figgatt, Dmitri Maslov, Kevin~A Landsman, Norbert~M Linke, Shantanu
  Debnath, and Christofer Monroe.
\newblock Complete 3-qubit grover search on a programmable quantum computer.
\newblock {\em Nature communications}, 8(1):1--9, 2017.

\bibitem{szablowski2021understanding}
Pawe{\l}~J Szab{\l}owski.
\newblock Understanding mathematics of grover’s algorithm.
\newblock {\em Quantum Information Processing}, 20(5):1--21, 2021.

\bibitem{grover1997quantum}
Lov~K Grover.
\newblock Quantum mechanics helps in searching for a needle in a haystack.
\newblock {\em Physical review letters}, 79(2):325, 1997.

\bibitem{https://doi.org/10.48550/arxiv.2207.05665}
Bin Yan and Nikolai~A. Sinitsyn.
\newblock An adiabatic oracle for grover's algorithm, 2022.

\bibitem{ibm}
IBM Quantum Experience \url{https://quantum-computing.ibm.com}.
\newblock [Accessed: 11-November-2021].

\bibitem{terhal2015quantum}
Barbara~M Terhal.
\newblock Quantum error correction for quantum memories.
\newblock {\em Reviews of Modern Physics}, 87(2):307, 2015.

\bibitem{preskill2018quantum}
John Preskill.
\newblock Quantum computing in the nisq era and beyond.
\newblock {\em Quantum}, 2:79, 2018.

\bibitem{clerk2010introduction}
Aashish~A Clerk, Michel~H Devoret, Steven~M Girvin, Florian Marquardt, and
  Robert~J Schoelkopf.
\newblock Introduction to quantum noise, measurement, and amplification.
\newblock {\em Reviews of Modern Physics}, 82(2):1155, 2010.

\bibitem{youssef2020measuring}
Rahaf Youssef.
\newblock Measuring and simulating t1 and t2 for qubits.
\newblock Technical report, Fermi National Accelerator Lab.(FNAL), Batavia, IL
  (United States), 2020.

\bibitem{rost2020simulation}
Brian Rost, Barbara Jones, Mariya Vyushkova, Aaila Ali, Charlotte Cullip,
  Alexander Vyushkov, and Jarek Nabrzyski.
\newblock Simulation of thermal relaxation in spin chemistry systems on a
  quantum computer using inherent qubit decoherence.
\newblock {\em arXiv preprint arXiv:2001.00794}, 2020.

\bibitem{L__2007}
C.~Lü, J.L. Cheng, M.W. Wu, and I.C. da~Cunha~Lima.
\newblock Spin relaxation time, spin dephasing time and ensemble spin dephasing
  time in n-type {GaAs} quantum wells.
\newblock {\em Physics Letters A}, 365(5-6):501--504, jun 2007.

\bibitem{skinner1986pure}
JL~Skinner and D~Hsu.
\newblock Pure dephasing of a two-level system.
\newblock {\em The Journal of Physical Chemistry}, 90(21):4931--4938, 1986.

\bibitem{ferraro2019phonon}
E~Ferraro, M~Fanciulli, and M~De~Michielis.
\newblock Phonon-induced relaxation and decoherence times of the hybrid qubit
  in silicon quantum dots.
\newblock {\em Physical Review B}, 100(3):035310, 2019.

\bibitem{Dudley:10}
Angela Dudley, Michael Nock, Thomas Konrad, Filippus~S. Roux, and Andrew
  Forbes.
\newblock Amplitude damping of laguerre-gaussian modes.
\newblock {\em Opt. Express}, 18(22):22789--22795, Oct 2010.

\end{thebibliography}
\end{document}